\documentclass[aps,pra,
   amsfonts,
   amssymb,
   amsmath,
   reprint,
   superscriptaddress,
   eqsecnum,
   showpacs,
   preprintnumbers,
   ]{revtex4-1}
\usepackage[T1]{fontenc}          
\usepackage{amsfonts}
\usepackage{amssymb}
\usepackage{amsmath}
\usepackage{amsthm}
\usepackage{amscd}
\usepackage{color}
\usepackage{graphicx}
\usepackage{epsfig}
\usepackage{subfigure}
\usepackage{fancyhdr}
\usepackage{dcolumn}
\usepackage{time}
\usepackage{hyperref}
\hypersetup{
    unicode=false,          
    pdftoolbar=true,        
    pdfmenubar=true,        
    pdffitwindow=false,     
    pdfstartview={FitH},    
    pdftitle={Paper},       
    pdfauthor={John Dawson},     
    pdfsubject={Phys Rev article},   
    pdfcreator={John Dawson},   
    pdfproducer={John Dawson}, 
    pdfkeywords={keyword1} {key2} {key3}, 
    pdfnewwindow=true,      
    colorlinks=true,        
    linkcolor=red,          
    citecolor=blue,         
    filecolor=magenta,      
    urlcolor=cyan           
}
\definecolor{dark}{gray}{.5}
\definecolor{light}{gray}{.75}
\definecolor{darkmagenta}{rgb}{.5,0,.5}
\theoremstyle{definition}

\theoremstyle{remark}

\newcommand{\rD}{\text{D}}

\newcommand{\bnabla}{\boldsymbol{\nabla}}

\newcommand{\bP}{\mathbf{P}}

\newcommand{\bj}{\mathbf{j}}
\newcommand{\bk}{\mathbf{k}}

\newcommand{\bq}{\mathbf{q}}
\newcommand{\br}{\mathbf{r}}

\newcommand{\bv}{\mathbf{v}}

\newcommand{\bx}{\mathbf{x}}
\newcommand{\by}{\mathbf{y}}

\newcommand{\calG}{\mathcal{G}}
\newcommand{\calH}{\mathcal{H}}

\newcommand{\calL}{\mathcal{L}}

\newcommand{\calN}{\mathcal{N}}

\newcommand{\hH}{\hat{H}}

\newcommand{\hN}{\hat{N}}

\newcommand{\hrho}{\hat{\rho}}
\newcommand{\kBolt}{k_{\text{B}}}              
\newcommand{\QED}%
   {$\mathcal{Q\kern-.1em \lower.6ex\hbox{$\mathcal{E}$}\kern-.1667em D}$}
\newcommand{\QCD}%
   {$\mathcal{Q\kern-.1em \lower.6ex\hbox{$\mathcal{C}$}\kern-.1667em D}$}

\newcommand{\SUSYext}%
   {$\mathcal{S \kern-0.08em \lower 0.5ex \hbox{$\mathcal{U}$}
   \kern-0.05em S \kern-0.2em \lower 0.5ex
   \hbox{$\mathcal{Y}$}}\kern-0.05em{}_{\text{ext}}$}
\newcommand{\Quad}[1]{\quad\text{#1}\quad}         
\newcolumntype{d}[1]{D{.}{.}{#1}}
\newcommand{\Set}[1]{\bigl ( \, #1 \, \bigr )}     
\newcommand{\Diag}[1]{\mathrm{diag}( \, #1 \, )}   
\newcommand{\rd}{\mathrm{d}}
\newcommand{\Partial}[4]
   {\Bigl ( \frac{\partial #1 }{\partial #2 } \Bigr )_{\! #3, #4 }}
\newcommand{\Intk}{\int\!\!%
   \frac{ \mathrm{d}^3 k }{ (2\pi)^3 } }           

\newcommand{\Intq}{\int\!\!%
   \frac{ \mathrm{d}^3 q }{ (2\pi)^3 } }           
\newcommand{\Inttau}{\int_{0}^{\beta}\!\!\! \mathrm{d} \tau} 
\newcommand{\Det}[1]{\det [ \, #1 \, ]}

\newcommand{\Exp}[1]{\exp \{ \, #1 \, \} }

\newcommand{\Ln}[1]{\ln [ \, #1 \, ]}

\newcommand{\Tr}[1]{\mathrm{Tr} [ \, #1 \, ]}

\newcommand{\bra}[1]%
   {\ensuremath{\langle \, #1 \, |}}
\newcommand{\Bra}[1]%
   {\ensuremath{\langle \, #1 \, |}}
\newcommand{\bigbra}[1]%
   {\ensuremath{\Bigl \langle \, #1 \, \Bigr |}}
\newcommand{\ket}[1]%
   {\ensuremath{| \, #1 \, \rangle}}
\newcommand{\Ket}[1]%
   {\ensuremath{| \, #1 \, \rangle}}
\newcommand{\bigket}[1]%
   {\ensuremath{\Bigl | \, #1 \, \Bigr \rangle}}
\newcommand{\braket}[2]%
   {\ensuremath{\langle \, #1 \, | \, #2 \, \rangle}}
\newcommand{\Braket}[2]%
   {\ensuremath{\langle \, #1 \, | \, #2 \, \rangle}}
\newcommand{\matrixelement}[3]%
   {\ensuremath{\langle \, #1 \, | \, #2 \, | \, #3 \, \rangle}}
\newcommand{\MatEl}[3]%
   {\ensuremath{\langle \, #1 \, | \, #2 \, | \, #3 \, \rangle}}
\newcommand{\pbra}[1]%
   {\ensuremath{( \, #1 \, |}}
\newcommand{\pket}[1]%
   {\ensuremath{| \, #1 \, )}}    
\newcommand{\pbraket}[2]%
   {\ensuremath{( \, #1 \, | \, #2 \, )}}
\newcommand{\braV}[1]%
   {\ensuremath{\langle \, #1 \, \Vert}}
\newcommand{\ketV}[1]%
   {\ensuremath{\Vert \, #1 \, \rangle}}
\newcommand{\Comm}[2]%
   {\ensuremath{[ \, #1, #2 \, ]}}
\newcommand{\AntiComm}[2]%
   {\ensuremath{\{ \, #1, #2 \, \}}}
\newcommand{\Pbracket}[2]%
   {\ensuremath{\{ \, #1, #2 \, \} }}
\newcommand{\PBracket}[2]%
   { \ensuremath{ \{ \, #1, #2 \, \}_{\lower1.0ex\hbox{\scriptsize \text{PB}}} } }
\newcommand{\wedgeComm}[2]
   {\ensuremath{[ \, #1, #2 \, ]_{\lower1.0ex\hbox{\scriptsize $\wedge$}} }}
\newcommand{\Expect}[1]
   {\ensuremath{\langle \, #1 \,  \rangle}}
\newcommand{\expect}[1]%
   {\ensuremath{\langle \, #1 \,  \rangle}}
\newcommand{\Expectbig}[1]%
   {\ensuremath{\Bigl \langle \, #1 \, \Bigr \rangle}}
\newcommand{\expectbig}[1]%
   {\ensuremath{\Bigl \langle \, #1 \, \Bigr \rangle}}
\newcommand{\expectc}[2]%
   {\ensuremath{\langle \, \{ \, #1 , #2 \, \} \, \rangle}}
\newcommand{\expectq}[2]%
   {\ensuremath{\langle \, [ \, #1 , #2 \, ] \, \rangle}}
\newcommand{\expectT}[1]%
   {\ensuremath{\langle \, \mathcal{T} \{ \, #1 \, \} \, \rangle}}
\newcommand{\expectaT}[1]%
   {\ensuremath{\langle \, \mathcal{T}^{\ast} \{ \, #1 \, \} \, \rangle}}
\newcommand{\expectTbig}[1]%
   {\ensuremath{\biggl \langle \, \mathcal{T}  \biggl \{ \, #1 \,%
\biggr \} \, \biggr \rangle}}
\newcommand{\expectTabig}[1]%
   {\ensuremath{\biggl \langle \, \mathcal{T}^{\ast} \biggl \{ \, #1 \,%
\biggr \} \, \biggr \rangle}}
\newcommand{\Tproduct}[1]%
   {\ensuremath{\mathcal{T} \{ \, #1 \, \} } }
\newcommand{\aTproduct}[1]%
   {\ensuremath{\mathcal{T}^{\ast} \{ \, #1 \, \} } }
\newcommand{\Nproduct}[1]%
   {\ensuremath{\mathcal{N} \{ \, #1 \, \} } }
\newcommand{\ctpTproduct}[1]%
   {\ensuremath{\mathcal{T}_{\mathcal{C}} \{ \, #1 \, \} }}
\newcommand{\tauordered}[1]%
   {\ensuremath{\mathcal{T}_{\tau} \{ \, #1 \, \} }}
\newcommand{\expectTproduct}[1]%
   {\ensuremath{\langle \, \mathcal{T} \{ \, #1 \, \} \, \rangle}}
\newcommand{\expectTCproduct}[1]%
   {\ensuremath{\langle \, \mathcal{T}_{\mathcal{C} \{ \, #1 \, \} \, \rangle}}}
\newcommand{\expectComm}[2]%
   {\ensuremath{\langle \, [ \, #1 , #2 \, ] \, \rangle}}
\newcommand{\expectPbracket}[2]%
   {\ensuremath{\langle \, \{ \, #1 , #2 \, \} \, \rangle}}
\newcommand{\threej}[6]%
{\begin{pmatrix} #1 & #2 & #3 \\ #4 & #5 & #6 \end{pmatrix}}
\newcommand{\sixj}[6]%
{\begin{Bmatrix} #1 & #2 & #3 \\ #4 & #5 & #6 \end{Bmatrix}}
\newcommand{\ninej}[9]%
{\begin{Bmatrix} #1 & #2 & #3 \\ #4 & #5 & #6 \\%
 #7 & #8 & #9 \end{Bmatrix}}
\newcommand{\reducedme}[3]%
{\langle \, #1 \, \Vert \, #2 \, \Vert \, #3 \, \rangle }
\begin{document}
%
%
%
%
\title[Superfluidity]
   {The Josephson relation for the superfluid density and \\
   the connection to the Goldstone theorem in dilute Bose atomic gasses}

\author{John F. Dawson}
\email{john.dawson@unh.edu}
\affiliation{Department of Physics,
   University of New Hampshire,
   Durham, NH 03824}

\author{Bogdan~Mihaila}
\email{bmihaila@lanl.gov}
\affiliation{Materials Science and Technology Division,
   Los Alamos National Laboratory,
   Los Alamos, NM 87545}

\author{Fred Cooper}
\email{cooper@santafe.edu}
\affiliation{Santa Fe Institute,
   Santa Fe, NM 87501}
\affiliation{Theoretical Division,
   Los Alamos National Laboratory,
   Los Alamos, NM 87545}


\begin{abstract}
We derive the Josephson relation for a dilute Bose gas in the framework of an auxiliary-field resummation of the theory in terms of the normal- and anomalous-density condensates. The mean-field phase diagram of this theory features two critical temperatures, $T_c$ and $T^\star$, associated with the presence in the system of the Bose-Einstein condensate and superfluid state, respectively. In this context, the Josephson relation shows that the  superfluid density is related to a second order parameter, the square of the anomalous-density condensate. This is in contrast with the corresponding result in the Bose gas theory without an anomalous condensate, which predicts that the superfluid density is proportional to the BEC condensate density. Our findings are consistent with the prediction that in the temperature range between $T_c$ and $T^\star$ a fraction of the system is in the superfluid state in the absence of the BEC condensate. This situation is similar to the case of dilute Fermi gases, where the superfluid density is proportional to the square of the gap parameter.
The Josephson relation relies on the existence of zero energy and momentum excitations showing the intimate relationship between superfluidity and the Goldstone theorem.  
\end{abstract}
%
%
%
%
\maketitle
%
%
\section{Introduction}

Superfluidity represents a fundamental property of matter that causes liquids to lose all their resistance to internal movement~\cite{r:London:1950fk}. 
In the limit of dilute Bose gases, Josephson showed~ that the superfluid density\cite{r:Josephson:1966fk} is proportional to the condensate density given by the square of the broken-symmetry order parameter, $\phi$. The Josephson relation reads~\cite{r:Josephson:1966fk,r:Baym:1967uq}
\begin{equation}
    \rho_s = - \frac{m^2 \phi^2}{\hbar^2} \lim_{q \rightarrow 0}  \frac{1}{ q^2  \, \tilde G_{11}(q,0)} \>,
    \label{eq:JR}
\end{equation}
where  $\tilde G_{11}(k,0)$ is the Fourier transform of the connected Green function corresponding to the expectation value $\Expect{\phi^\star(x) \phi(x')}$, evaluated at zero energy transfer and momentum transfer, $q$. 
The Josephson relation was re-derived in the framework of  diagrammatic perturbation theory by Holzmann and Baym~\cite{PhysRevB.76.092502}. In both approaches, the Josephson relation implies that the Bose-Einstein condensate (BEC) and the superfluid state are intimately linked. Hence, the critical temperature, $T_c$, corresponding to the temperature where the BEC condensate first appears in the system, coincides also with the temperature where superfluidity sets in. Unfortunately, it is well-known that dilute Bose gases theories involving a single order parameter predict a first-order phase transition, contrary to the second-order transition expected for the universality class of the Bose gas~\cite{PhysRevD.15.2897}. 

Recently, we reformulated the theory of dilute Bose gases by treating the normal and anomalous densities on equal footing~\cite{PhysRevLett.105.240402}. In the leading-order approximation, our auxiliary-field loop expansion predicts a second-order BEC phase transition. The leading-order auxiliary-field (LOAF) approximation is a conserving and gapless approximation with a non-perturbative character and can be applied outside the regime of weakly-interacting particles, while still preserving the correct weak-coupling limits~\cite{PhysRevA.84.023603}. LOAF produces the same slope of the linear departure of the critical temperature from the noninteracting limit derived by Baym \emph{et al.}~\cite{r:Baym:2000fk} using a large-N expansion for the critical theory. Unlike the large-N expansions developed by Baym \emph{et al.}~\cite{r:Baym:2000fk}, the LOAF approximation can be used at all temperatures.

Unlike all other theories of interacting dilute Bose gases, however, the LOAF phase diagram features two critical temperatures: the critical temperature~$T_c$, where the BEC condensate appears first, and a temperature~$T^\star > T_c$, where the anomalous-density condensate,~$A$, turns on. We used the Landau two-fluid phenomenological model~\cite{r:Landau:1941ly,*r:Landau:1947} to show~\cite{Anomalon:2011} that in LOAF a superfluid state is present below $T^\star$ and that the superfluid density is proportional to $A^2$~\cite{Anomalon:2011}. Moreover, LOAF indicates that in the temperature range between $T_c$ and $T^\star$ the system supports zero energy and momentum excitations, which correspond to composite-field Goldstone states. This is in addition to the usual Goldstone theorem for $T < T_c$, where the natural U(1) symmetry breaking of the theory leads to the usual BEC condensate as the Goldstone state below $T_c$.
Introducing a hypothetical U(1) gauge vector meson into the system, we were able to show that the mass generated through the Anderson-Higgs mechanism~\cite{PhysRev.117.648,*PhysRev.130.439,*PhysRevLett.13.321,*PhysRevLett.13.585,*r:Higgs:1964zr,*PhysRevLett.13.508} can be related to the superfluid density via the Meissner effect, which also gives that $\rho_s$ is proportional to $A^2$. Hence, LOAF predicts that part of the system is in a superfluid state, in the absence of a BEC condensate, in the temperature range between the BEC critical temperature and the temperature where superfluidity sets in. According to Eq.~\eqref{eq:JR}, the LOAF approximation appears to contradict the Josephson relation for an interacting dilute Bose gas.

In this paper, we will derive the form of the Josephson relation in the LOAF model. We will show that in our auxiliary-field formalism, the superfluid density is indeed proportional to the square of the anomalous-density condensate. The latter is identified as a second order parameter in LOAF. The Josephson relation for the superfluid density in LOAF reads
\begin{equation}
   \rho_s = 
       - \,
   \frac{8 m^2 \, A^2}{\hbar^2} \,
   \lim_{q \rightarrow 0}
   \frac{1}{q^2 \, \tilde{\calG}^{A}{}_A(q,0)} 
   \>.
    \label{eq:JR-loaf}
\end{equation}
Here, $\tilde \calG^{A}{}_A$ is the Fourier transform of the connected propagator corresponding to the expectation value $\Expect{A^\star A}$. Furthermore, we will study in detail the connection between the Josephson relation and the Goldstone theorem in the LOAF model.

The form of the Josephson relation described in Eq.~\eqref{eq:JR-loaf} is not surprising. In fact, the Josephson relation for the case of interacting dilute Fermi gases was derived by Taylor  and has the form~\cite{PhysRevB.77.144521}
\begin{equation}
    \rho_s = - \frac{m_B^2 \, \Delta^2}{\hbar^2} \lim_{q \rightarrow 0} \frac{1}{ q^2 \tilde D_{11}(q,0)} \>,
    \label{eq:JR-fermi}
\end{equation}
where $\tilde D_{11}$ is the Fourier transform of the connected propagator corresponding to the expectation value $\Expect{\Delta^\star \Delta}$ for quasi-particle bosons of mass, $m_B$, constructed as pairs of mass $m_F$ fermions ($m_B = 2 m_F$). Given that the gap parameter, $\Delta$, is in effect an auxiliary field, $\Delta = \Expect{\psi_\uparrow \psi_\downarrow}$, the propagator $D_{11}$ is an auxiliary-field propagator, similarly to what is described in Eq,~\eqref{eq:JR-loaf}.  It is important to note here that the fermionic counterpart of the LOAF model for dilute Bose gases we discuss in this paper~\cite{PhysRevA.Fermi:2011}, is precisely the standard Bardeen-Cooper-Schrieffer  (BCS) ansatz \cite{PhysRevLett.71.3202,PhysRevB.55.15153} that underlines the derivation of Eq.~\eqref{eq:JR-fermi}. The fact that in LOAF the Josephson relation \eqref{eq:JR-loaf} is different from the classical expression \eqref{eq:JR} is due to the fact that in our auxiliary field formalism, we treat the normal and anomalous density condensate condensates on equal footing.  In this sense, the auxiliary-field formalism discussed here provides a unified approach to the study of fermionic and bosonic atom gases. The detailed derivation of the LOAF approximation for the case of dilute Bose and Fermi atomic gases can be found elsewhere~\cite{PhysRevA.Fermi:2011,PhysRevA.Bose:2011}.

This paper is organized as follows: We begin by reviewing the calculation of thermodynamic variables in the auxiliary-field formalism in Sec.~\ref{J.s:thermo}.  In Sec.~\ref{G.s:goldstone} we discuss the Goldstone theorem in the auxiliary-field formalism, and we show that Goldstone modes are to be expected both for temperatures below $T_c$, but also in the temperature range $T_c < T < T^\star$. Turning our attention to the study of the superfluid phase, we begin by deriving the superfluid density in the Landau two-fluid model, and we show explicitly that in LOAF the superfluid density is proportional the square of the anomalous-density condensate in Sec.~\ref{s:landaucalc}. In Sec.~\ref{s:microcalc}, we calculate superfluid density in our microscopic theory and show the connection with the current-current correlation function. Here we follow an approach similar to that discussed by Taylor \emph{et al.} in the context of interacting Fermi gases~\cite{PhysRevA.74.063626}.  Finally, in Sec.~\ref{J.m.s:twist} we derive the Josephson relation for superfluidity in our auxiliary-field formalism and show that the square of the anomalous-density condensate is the second order parameter in LOAF.

%
%
\section{\label{J.s:thermo}Thermodynamics of the Bose system in the auxiliary field formalism}

We use here a grand canonical ensemble to describe the system in equilibrium.  In this case, the normalized thermodynamic probability density matrix $\hrho$ for the grand canonical ensemble is obtained by minimizing the entropy, defined by $S/\kBolt = - \Tr{ \hrho \Ln{ \hrho }}$, such that the average energy $E = \Tr{ \hrho \hH}$ and average number of particles $N = \Tr{ \hrho \hN }$ are held constant.  The solution for the minimization is given by
\begin{align}
   \hrho
   &=
   \frac{1}{Z[T,\mu,V]} \, e^{- \beta ( \, \hH - \mu \, \hN \, ) } \>,
   \label{thm.e:rhoGCE} \\
   Z[T,\mu,V]
   &=
   e^{- \beta \, \Omega[T,\mu,V] }
   =
   \Tr{ e^{- \beta \, ( \, \hH - \mu \, \hN \, ) } } \>.
   \notag
\end{align}
where we have set $\beta = 1 / ( \kBolt T )$ and $\Omega[T,\mu,V]$ is the grand potential.
In our auxiliary field formalism, we will compute the thermodynamic grand potential using a path integral formalism.  Once we have the potential, we can find all the thermodynamical properties of the system.  For instance, 
using the second law of thermodynamics, $T \, \rd S[E,N,V] = \rd E - \mu \, \rd N + p \, \rd V$, we find 
\begin{subequations}\label{thm.e:SNp}
\begin{align}
   S[T,\mu,V]
   &=
   - \Partial{\,\Omega}{T}{\mu}{V} \>,
   \label{thm.e:S} \\
   N[T,\mu,V]
   &=
   - \Partial{\,\Omega}{\mu}{T}{V} \>,
   \label{thm.e:N}\\
   p[T,\mu,V]
   &=
   - \Partial{\,\Omega}{V}{T}{\mu} \>.
   \label{thm.e:p}
\end{align}
\end{subequations}
From the second law, the energy is given by
\begin{equation}\label{thm.e:Evalue}
   E
   =
   \Omega + T S + \mu N \>.
\end{equation}

In the following, we review the auxiliary field formalism for dilute Bose gasses \cite{PhysRevLett.105.240402,PhysRevA.Bose:2011} using the imaginary time formalism.

%
%
\subsection{Path integrals}

The partition function is given by a path integral \cite{r:Negele:1988fk},
\begin{equation}\label{J.pf.e:Zdef}
   Z[V,\mu,\beta]
   =
   \calN
   \iint \rD \phi \, \rD \phi^{\ast} \,
   e^{ - S[\phi,\phi^{\ast};V,\mu,\beta] } \>,
\end{equation}
where the Euclidian action $S[\phi,\phi^{\ast};V,\mu,\beta]$ is given by
\begin{equation}\label{J.pf.e:actiondef}
   S[\phi,\phi^{\ast};V,\mu,\beta]
   =
   \int [ \rd x ] \, \calL[\phi,\phi^{\ast};\mu] \>,
\end{equation}
and where we have introduced the notation,
\begin{equation}\label{J.pf.e:intdx}
   \int [ \rd x ]
   =
   \int \rd^3 x \Inttau \>.
\end{equation}
The Euclidian Lagrangian density is given by
\begin{align}
   &\calL[\phi,\phi^{\ast};\mu]
   \label{J.pf.e:L-I} \\
   & \quad
   =
   \frac{1}{2} \,
   \bigl [ \,
      \phi^{\ast}(x) \, \partial_{\tau} \, \phi(x)
      -
      \phi(x) \, \partial_{\tau} \, \phi^{\ast}(x) \,
   \bigr ]
   +
   \calH[\phi,\phi^{\ast};\mu] \>,
   \notag
\end{align}
where the Hamiltonian density is given by
\begin{equation}\label{J.pf.e:Hdef_i}
   \calH[\phi,\phi^{\ast};\mu]
   =
   \phi^{\ast}(x) \,
   \Bigl [ \,
      - \frac{ \hbar^2 \nabla^2}{2m} - \mu \,
   \Bigr ] \, 
   \phi(x)
   +
   \frac{\lambda}{2} \, | \phi(x) |^4 \>.
\end{equation}
Auxiliary fields $\chi(x)$ and $A(x)$ are introduced by means of the Hubbard-Stratonovitch transformation Lagrangian density, which takes the form
\begin{align}
   &\calL_{\text{aux}}[\phi,\phi^{\ast},\chi,A,A^{\ast}]
   =
   - \frac{1}{2 \lambda} \,
   \bigl [ \,
      \chi(x) - \lambda \, \sqrt{2} \, | \phi(x) |^2 \,
   \bigr ]^2
   \notag \\
   & \qquad
   +
   \frac{1}{2 \lambda} \,
   \bigl | \,
      A(x) 
      - 
      \lambda \, \phi^{2}(x) \,
   \bigr |^2
   \>,
   \label{J.pf.e:Laux}
\end{align}
which we add to Eq.~\eqref{J.pf.e:L-I}, giving a Euclidian action of the form,
\begin{align}
   &S[J,\Phi;V,\mu,\beta]
   \label{J.intro.e:S-II} \\
   & \quad
   =
   \frac{1}{2}
   \iint [\rd x] \, [\rd x'] \,
   \phi_{a}(x) \, G^{-1}{}^a{}_b(x,x') \, \phi^{b}(x)
   \notag \\
   & \qquad
   +
   \int [\rd x] \,
   \Bigl [ \,
      \frac{| A(x) |^2 - \chi^2(x)}{2 \lambda}
      -
      J_{\alpha}(x) \, \Phi^{\alpha}(x) \,
   \Bigr ] \>,
   \notag
\end{align}
where 
\begin{equation}\label{J.intro.e:Gvinvdef}
   G^{-1}{}^a{}_b(x,x')
   =
   \delta(x,x') \,
   \begin{pmatrix}
      h^{(+)} & -A(x) \\
      -A^{\ast}(x) & h^{(-)}
   \end{pmatrix} \>,
\end{equation}
with
\begin{gather}
   h^{(+)}
   =
   h + \partial_{\tau} \>,
   \qquad
   h^{(-)}
   =
   h - \partial_{\tau} \>,
   \label{J.intro.e:hplusminus} \\
   h
   =
   -
   \frac{ \hbar^2 \nabla^2}{2m}
   +
   \sqrt{2} \, \chi(x)
   -
   \mu \>,
   \notag    
\end{gather}
and
\begin{align}
   &J_{\alpha}(x) \, \Phi^{\alpha}(x)
   =
   j^{\ast}(x) \phi(x) + j(x) \phi^{\ast}(x)
   \label{J.pf.e:JPhisum} \\
   & \qquad
   +
   s(x) \, \chi(x) + S^{\ast}(x) \, A(x) + S(x) \, A(x) \>.
   \notag
\end{align}

Here we have added currents and introduced a two-component notation using Roman indices $a,b,c,\dotsb$ for the fields $\phi(x)$ and $\phi^{\ast}(x)$ and currents $j(x)$ and $j^{\ast}(x)$,
\begin{subequations}\label{J.pf.e:phijnotes}
\begin{align}
   \phi^a(x)
   &=
   \Set{ \phi(x), \phi^{\ast}(x) } \>,
   \label{J.pf.e:phinote} \\
   \phi_a(x)
   &=
   \Set{ \phi^{\ast}(x), \phi(x) } \>,
   \notag \\
   j^a(x)
   &=
   \Set{ j(x), j^{\ast}(x) } \>,
   \label{J.pf.e:jnote} \\
   j_a(x)
   &=
   \Set{ j^{\ast}(x), j(x) } \>,
   \notag
\end{align}
\end{subequations}
for $a=1,2$, and a three-component notation using Roman indices $i,j,k,\dotsb$ for the fields $\chi(x)$, $A(x)$, and $A^{\ast}(x)$,
\begin{subequations}\label{J.pf.e:chiSnots}
\begin{align}
   \chi^{i}(x)
   &=
   \Set{ \chi(x), A(x), A^{\ast}(x) } \>,
   \label{J.pf.e:chinotes} \\
   \chi_{i}(x)
   &=
   \Set{ \chi(x), A^{\ast}(x), A(x) } \>,
   \notag \\
   S^{i}(x)
   &=
   \Set{ s(x), S(x), S^{\ast}(x) } \>,
   \label{J.pf.e:Snotes} \\
   S_{i}(x)
   &=
   \Set{ s(x), S^{\ast}(x), S(x) } \>,
   \notag
\end{align}
\end{subequations}
for $i=0,1,2$.   For convenience, we also define five-component fields with Greek indices $\Phi^{\alpha}(x)$ and currents $J^{\alpha}(x)$,
\begin{subequations}\label{J.intro.e:PhiJdefs-I}
\begin{align}
   \Phi^{\alpha} 
   &= 
   \Set{\phi^a(x),\chi^i(x)}
   =
   \Set{\phi,\phi^{\ast},\chi,A,A^{\ast}} \>,
   \label{J.e:Phidefs-I} \\
   \Phi_{\alpha} 
   &= 
   \Set{\phi_a(x),\chi_i(x)}
   =
   \Set{\phi^{\ast},\phi,\chi,A^{\ast},A} \>,
   \notag \\
   J^{\alpha}(x)
   &=
   \Set{j^a(x),S^i(x) }
   =
   \Set{ j,j^{\ast},s,S,S^{\ast}} \>,
   \label{J.intro.e:Jdefs-I} \\
   J_{\alpha}(x)
   &=
   \Set{j_a(x),S_i(x) }
   =
   \Set{ j^{\ast},j,s,S^{\ast},S} \>,
   \notag   
\end{align}
\end{subequations}
The metric which raises and lowers indices is given by
\begin{equation}\label{J.intro.e:metric}
   \eta^{\alpha\beta}
   =
   \eta_{\alpha\beta}
   =
   \begin{pmatrix}
      0 & 1 & 0 & 0 & 0 \\
      1 & 0 & 0 & 0 & 0 \\
      0 & 0 & 1 & 0 & 0 \\
      0 & 0 & 0 & 0 & 1 \\
      0 & 0 & 0 & 1 & 0
   \end{pmatrix} \>.
\end{equation}
The partition function is now a functional of the currents and given by a path integral over all fields, which we write symbolically as
\begin{align}
   Z[J;V,\mu,\beta]
   &=
   e^{ - W[J;V,\mu,\beta] }
   \label{J.pf.e:Zdef-II} \\
   &=
   \calN\!
   \int \! \rD \Phi \,
   e^{ - S[\Phi,J;V,\mu,\beta] } \>,
   \notag
\end{align}
with $S[\Phi,J;V,\mu,\beta]$ now given by Eq.~\eqref{J.intro.e:S-II}.  
The thermodynamic average value of the fields are given by
\begin{equation}\label{J.pf.e:dZdJ}
   \Expect{ \Phi^{\alpha}(x) }
   =
   \frac{1}{Z} \frac{\delta Z[J]}{\delta J_{\alpha}(x)} \Big |_{J=0}
   =
   - \frac{\delta W[J]}{\delta J_{\alpha}(x)} \Big |_{J=0} \>,
\end{equation}
evaluated at zero currents, and the connected two-point functions are given by the Green function
\begin{equation}\label{J.pi.e:Green}
   \calG_{\alpha\beta}(x,x')
   =
   - \frac{\delta^2 W[J]}{\delta J_{\alpha}(x) \, \delta J_{\beta}(x')} \Big |_{J=0} \>.
\end{equation}

%
%
\subsection{\label{ss:effaction}Effective action}

After introducing the auxiliary fields, the action is now quadratic in the $\phi$-fields and can be integrated out.  The partition function is then given by
\begin{equation}\label{J.ea.e:Zdef-III}
   Z[J]
   =
   e^{ -W[J] }
   =
   \calN\!
   \int \! \rD \chi \,
   e^{ - S_{\text{eff}}[\chi,J] / \epsilon } \>,
\end{equation}
where $W[J] \equiv \beta \, \Omega[J]$ and where
\begin{align}
   &S_{\text{eff}}[\chi,J] 
   \label{J.ea.e:Seff-I} \\
   & \quad
   =
   - \frac{1}{2}
   \iint [\rd x] \, [\rd x'] \,
   j_{a}(x) \, G^a{}_b[\chi](x,x') \, j^{b}(x')
   \notag \\
   & \qquad
   +
   \int [\rd x] \,
   \Bigl [ \,
      \frac{| A(x) |^2 - \chi^2(x)}{2 \lambda}
      -
      S_{i}(x) \, \chi^{i}(x) \,
      \notag \\
      & \qquad\qquad\qquad
      +
      \frac{1}{2}  
      \Tr{ \Ln{ G^{-1}[\chi](x,x) } } \,
   \Bigr ] \>.
   \notag
\end{align}
The dimensionless parameter $\epsilon$ in Eq.~\eqref{J.ea.e:Zdef-III} allows us to count loops for the auxiliary-field propagators in the effective action. 

Next we expand the effective about the stationary points $\chi_0^{i}(x)$, defined by $\delta S_{\text{eff}}[ \chi,J ] /  \delta \chi_{i}(x) = 0$, i.e
\begin{align}
   \frac{\chi_0(x)}{\lambda}
   &=
   \sqrt{2} \,
   \bigl \{ \,
      | \phi_0(x) |^2
      +
      \Tr{ G[\chi_0](x,x) } / 2  \,
   \bigr \} 
   - 
   s(x)
  \notag \\
  \frac{A_0(x)}{\lambda}
   &=
   \phi^2_0(x) 
   +
   G^{2}{}_{1}[\chi_0](x,x)
   + 
   2 \, S(x) \>,
   \label{J.ea.e:chi0A0eqs}
\end{align}
where $\phi^a_0(x)$ is given by
\begin{equation}\label{J.ea.e:phi0def}
   \phi^a_0[\chi_0,J](x)
   =
   \varphi^a_0[\chi_0](x)
   +
   \int [\rd x'] \, G[\chi_0]^a{}_b(x,x') \, j^b(x') \>,
\end{equation}
where $\varphi^a_0[\chi_0](x)$ is a solution of the homogenous equation,
\begin{equation}\label{J.ea.e:varphieq}
   G^{-1}{}^a{}_b[\chi_0](x) \, \varphi^b_0[\chi_0](x)
   =
   0 \>.
\end{equation}
Eqs.~\eqref{J.ea.e:chi0A0eqs} are called the ``gap'' equations.
The fields $\chi^i_0[J](x)$ at the stationary points are functionals of all the currents $J^{\alpha}(x)$.  
Expanding the effective action about the stationary point, we find
\begin{align}\label{J.ea.e:Seffexpand}
   S_{\text{eff}}[ \chi,J ]
   &=
   S_{\text{eff}}[ \chi_0,J ]
   \\ & \qquad
   +
   \frac{1}{2} \iint [\rd x] \, [\rd x'] \,
   D^{-1}_{ij}[\chi_0](x,x')
   \notag \\
   & \qquad 
   \times
   ( \chi^i(x) - \chi^i_0(x) ) \,
   ( \chi^j(x') - \chi^j_0(x') )
   +
   \dotsb
   \notag
\end{align}
where $D_{ij}^{-1}[\chi_0](x,x')$ is given by the second-order derivatives
\begin{align}
   D^{-1}_{ij}[\chi_0](x,x'))
   &=
   \frac{ \delta^2 \, S_{\text{eff}}[ \chi^a] }
        { \delta \chi^i(x) \, \delta \chi^j(x') } \, \bigg |_{\chi_0} 
   \label{J.ea.e:Dinverse} \\
   &=
   \frac{1}{\lambda} \tilde{\eta}_{ij} \, \delta(x,x')
   +
   \Pi_{ij}[\chi_0](x,x') \>,
   \notag
\end{align}
evaluated at the stationary points.  Here
\begin{equation}\label{J.ea.e:tildeetadef}
   \tilde{\eta}_{ij}
   =
   \begin{pmatrix}
      -1 & 0 & 0 \\
      0 & 0 & 1/2 \\
      0 & 1/2 & 0
   \end{pmatrix} \>,
\end{equation}
and $\Pi_{ij}[\chi_0](x,x')$ is the polarization, given symbolically by
\begin{align}
   \Pi^{ij}[\chi_0]
   &=
   \frac{1}{2}
   \Bigl \{ \,
      - 
      \phi_0[\chi_0] \circ V^{ij}[\chi_0] \circ \phi_0[\chi_0]
      \label{J.ea.e:Pidef} \\
      & \qquad\quad
      +
      \Tr{ G[\chi_0] \circ V^{i} \circ G[\chi_0] \circ V^{j} } \,
   \Bigr \} \>,
   \notag
\end{align}
with
\begin{align}
   V^{i}
   &=
   \frac{\delta G^{-1}[\chi]}{\delta \chi_{i}} \>,
   \label{J.ea.e:VVdefs} \\
   V^{ij}[\chi_0]
   &=
   V^{i} \circ G[\chi_0] \circ V^{j}
   +
   V^{j} \circ G[\chi_0] \circ V^{i} \>.
   \notag
\end{align}

We perform the remaining gaussian path integral over the fields $\chi_i$, obtaining the result for the grand potential
\begin{align}
   \epsilon \, W[J]
   &=
   S_0
   +
   S_{\text{eff}} [\chi_0,J] 
   \label{J.ea.e:WtoSD} \\
   & \qquad
   +
   \frac{\epsilon}{2} 
   \int \! [\rd x] \, \Tr{ \Ln{ D^{-1}[\chi_0,J](x,x) } }
   +
   \dotsb  \>,
   \notag
\end{align}
where $S_0$ is a normalization constant.  The fields are given by the expansion,
\begin{align}
   \epsilon \, \Phi^{\alpha}[J](x)
   &=
   - \epsilon \, \frac{\delta W[J] }{\delta J_{\alpha}(x)}
   \label{J.ea.e:Phiepsilon} \\
   &=
   \Phi^{(0)\,\alpha}[J](x)
   +
   \epsilon \, \Phi^{(1) \,\alpha}[J](x)
   +
   \dotsb \>.
   \notag
\end{align}
We calculate the order $\epsilon$ corrections to the fields from \eqref{J.ea.e:Phiepsilon}, evaluated at zero currents.  We work these out explicitly for each field to zeroth order and find
\begin{subequations}\label{J.ea.e:Phi0order}
\begin{align}
   \phi^{(0)\,a}(x)
   &=
   \varphi^a_0[\chi_0](x) \>,
   \label{J.ea.e:phi0order} \\
   \chi^{(0)}(x)
   &=
   \chi_{0}(x) \>,
   \label{J.ea.e:chi0order} \\
   A^{(0)}(x)
   &=
   A_{0}(x) \>,
   \label{J.ea.e:A0order} 
\end{align}
\end{subequations}
where $\varphi^a_0[\chi_0](x)$ is given by Eq.~\eqref{J.ea.e:varphieq} and $\chi_{0}(x)$ and $A_{0}(x)$ are given by Eqs.~\eqref{J.ea.e:chi0A0eqs}, evaluated at $s(x) = S(x) = 0$.  Diagrams for the first order fields $\Phi^{(1) \,\alpha}[J](x)$ are given in Ref.~\onlinecite{PhysRevA.Bose:2011}.

The grand potential $\Gamma[\Phi]$ as a functional of the fields $\Phi$ (rather than the currents $J$) is constructed by a Legendre transformation, 
\begin{align}
   &\epsilon \, \Gamma[\Phi]
   =
   \epsilon \int [\rd x] \, J_{\alpha}(x) \, \Phi^{\alpha}(x)
   +
   \epsilon \, W[J]
   \label{J.ea.e:vertexfctdef} \\
   &=
   \Gamma_0
   +
   \frac{1}{2}\! \iint [\rd x] \, [\rd x'] \,
   \phi_a(x) \, G^{-1}[\chi]^{a}{}_{b}(x,x') \, \phi^b(x')
   \notag \\
   & 
   +
   \int\! [\rd x] \,
   \Bigl \{ \,
      \frac{ |A(x)|^2 - \chi^2(x) }{2\lambda}
      +
      \frac{1}{2} \,
      \Tr{ \Ln{ G^{-1}[\chi](x,x) } }
      \notag \\
      & \qquad
      +
      \frac{\epsilon}{2} \,
      \Tr{ \Ln{ D^{-1}[\chi_0,J](x,x) } }
    \Bigr \}
    +
    \dotsb
    \notag 
\end{align}
Here $\Gamma_0$ is a normalization constant.  Then the currents are given by
\begin{equation}\label{J.ea.e:dGammadphidchi}
   J_{\alpha}[\Phi](x)
   =
   \frac{\delta \Gamma[\Phi]}{\delta \Phi^{\alpha}(x)} \>,
\end{equation}
and the inverse Green functions by
\begin{equation}\label{J.eq.e:Ginvdefs}
   \calG_{\alpha\beta}^{-1}(x,x')
   =
   \frac{\delta^2 \Gamma[\Phi]}{\delta \Phi^{\alpha}(x) \, \delta \Phi^{\beta}(x')} \>.
\end{equation}

%
%
\subsection{\label{J.ss:LOAFmodel}Leading order auxiliary field (LOAF) approximation}

For uniform $\tau$-independent systems, $\Phi^{\alpha}(x) \equiv \Phi^{\alpha}$ are all constants.  Let us first define the effective potential $V_{\text{eff}}[\Phi]$ as the grand potential per unit four-volume as a functional of the fields.  Then from \eqref{J.ea.e:vertexfctdef} for uniform systems,
\begin{align}
   &V_{\text{eff}}[\Phi] 
   \equiv 
   \epsilon \, \frac{\Gamma[\Phi]}{\beta V}
   =
   V_0
   +
   \chi' \, | \phi |^2
   -
   \frac{1}{2} \, [ \, A \, \phi^{\ast\,2} + A^{\ast} \, \phi^2 \, ]
   \notag \\
   & 
   +
   \frac{ |A|^2 }{2\lambda}
   -
   \frac{ ( \chi' + \mu )^2 }{ 4 \lambda }
   +
   \frac{1}{2} \,
   \Tr{ \Ln{ G^{-1}[\chi](x,x) } }
   \label{J.LOAF.e:Gamma-II}
\end{align}
Here we have set $\chi' = \sqrt{2} \, \chi - \mu$.  Expansion of the inverse Green function in a Fourier series gives
\begin{align}
\frac{1}{2} \,
   &\Tr{ \Ln{ G^{-1}[\chi](x,x) } }
   \label{J.OAF.e:Ginvexp} \\
   & \qquad
   =
   \frac{1}{2 \beta}
   \Intk \sum_n \, \Ln{ \Det{ \tilde{G}^{-1}(\bk,n) } }
   \notag \\
   & \qquad
   =
   \Intk 
   \frac{1}{2 \beta} \sum_n
   \Ln{ \omega_k^2 + \omega_n^2 }
   \notag \\
   & \qquad 
   =
   \Intk
   \Bigl \{ \,
      \frac{\omega_k}{2}
      +
      \frac{1}{\beta}
      \Ln{ 1 - e^{-\beta \omega_k} } \,
   \Bigr \} \>,
   \notag
\end{align}
where
\begin{equation}\label{J.LOAF.e:omegakdef}
   \omega_k
   =
   \sqrt{ ( \epsilon_k + \chi')^2 - |A|^2 } \>,
   \qquad
   \epsilon_k = \frac{\hbar^2 k^2}{2m} \>.
\end{equation}
Inserting this result into \eqref{J.LOAF.e:Gamma-II} gives
\begin{align}
   &V_{\text{eff}}[\Phi]
   =
   V_0
   +
   \chi' \, | \phi |^2
   -
   \frac{1}{2} \, [ \, A \, \phi^{\ast\,2} + A^{\ast} \, \phi^2 \, ]
   \label{J.LOAF.e:Gamma-III} \\
   & 
   + \!
   \frac{ |A|^2 }{2\lambda}
   - \!
   \frac{ ( \chi' + \mu )^2 }{ 4 \lambda }
   + \!
   \Intk
   \Bigl \{ \,
      \frac{\omega_k}{2}
      +
      \frac{1}{\beta}
      \Ln{ 1 - e^{-\beta \omega_k} } \,
   \Bigr \} \>,
   \notag
\end{align}
However this expression is not finite.  Expanding $\omega_k$ about $k \rightarrow \infty$, we find
\begin{equation}\label{J.LOAF.e:expandomegak}
   \omega_k
   =
   \epsilon_k
   +
   \chi'
   -
   \frac{ |A|^2}{2 \epsilon_k}
   +
   \dotsb \>.
\end{equation}
These three terms are responsible for the divergences in the integral in Eq.~\eqref{J.LOAF.e:Gamma-III}.  Our method of regularization is to subtract these three terms from the integrand and replace the coupling constant, chemical potential, and the normalization constant by renormalized values, which gives
\begin{align}
   &V_{\text{eff}}[\Phi]
   =
   V_R
   +
   \chi' \, | \phi |^2
   -
   \frac{1}{2} \, [ \, A \, \phi^{\ast\,2} + A^{\ast} \, \phi^2 \, ]
   +
   \frac{ |A|^2 }{2\lambda_R}
   \notag \\
   & \quad 
   - 
   \frac{ ( \chi' + \mu_R )^2 }{ 4 \lambda_R }
   + 
   \Intk
   \Bigl \{ \,
      \frac{1}{2} \,
      \Bigl [ \,
         \omega_k
         -
         \epsilon_k
         -
         \chi'
         +
         \frac{ |A|^2}{2 \epsilon_k} \,
      \Bigr ]
      \notag \\
      & \quad
      +
      \frac{1}{\beta}
      \Ln{ 1 - e^{-\beta \omega_k} } \,
   \Bigr \} \>,
   \label{J.LOAF.e:Gamma-IV}
\end{align}
In dilute atomic gasses, the renormalized coupling constant is related to the $s$-wave scattering length, $a_0$, by $\lambda_R = 4\pi \hbar^2 \, a_0 / m$.  
The regularization method is discussed in Ref.~\onlinecite{PhysRevA.Bose:2011}.

Evaluating $V_{\text{eff}}[\Phi]$ at the minimum where
\begin{equation}\label{J.LOAF.e:dVdPhi}
   \frac{\partial V_{\text{eff}}[\Phi]}{\partial \Phi^{\alpha}}
   =
   0 \>,
\end{equation}
yields the three equations,
\begin{subequations}\label{J.LOAF.e:gapeq-I}
\begin{align}
   &\qquad\qquad
   \begin{pmatrix}
      \chi' & - A \\
      -A^{\ast} & \chi'
   \end{pmatrix}
   \begin{pmatrix}
      \phi \\ \phi^{\ast}
   \end{pmatrix}
   =
   0 \>,
   \label{J.LOAF.e:gapeq-I-a} \\
   \frac{\chi' + \mu_R}{2\lambda_R}
   &=
   | \phi |^2
   \qquad\qquad\qquad
   \label{J.LOAF.e:gapeq-I-b} \\
   & \quad
   +
   \Intk \, 
   \Bigl \{ \,
      \frac{\epsilon_k + \chi'}{2 \omega_k} \, [ \, 2 n(\beta\omega_k) + 1 \, ]
      -
      \frac{1}{2} \,
   \Bigr \} \>,
   \notag \\
   \frac{A}{\lambda_R}
   &=
   \phi^2
   \label{J.LOAF.e:gapeq-I-c} \\
   & \quad
   -
   A \Intk \, 
   \Bigl \{ \,
      \frac{1}{2\omega_k}  \, [ \, 2 n(\beta\omega_k) + 1 \, ] 
      -
      \frac{1}{2\epsilon_k} \,
   \Bigr \} \>,
   \notag 
\end{align}
\end{subequations}
where $n(x) = 1 / ( e^{x} - 1 )$ is the free particle Bose number distribution function.  The particle density $\rho$ is given by Eq.~\eqref{thm.e:N} where the effective potential is evaluated at the minimum of the potential,
\begin{equation}\label{J.LOAF.e:rho-I}
   \rho
   =
   -
   \Bigl \{ \,
      \frac{\partial V_{\text{eff}}}{\partial \mu_R}
      +
      \frac{\partial V_{\text{eff}}}{\partial \Psi^{\alpha}} \,
      \frac{\partial \Psi^{\alpha}}{\partial \mu_R} \,
   \Bigr \}
   =
   \frac{\chi' + \mu_R}{2\lambda} \>.
\end{equation}
The condensate density is defined to be $\rho_c = | \phi |^2$ at the minimum of the effective potential.  From \eqref{thm.e:p}, the pressure is the negative of the effective potential
\begin{align}
   p
   &= 
   -
   V_R
   -
   \chi' \, | \phi |^2
   +
   \frac{1}{2} \, [ \, A \, \phi^{\ast\,2} + A^{\ast} \, \phi^2 \, ]
   -
   \frac{ |A|^2 }{2\lambda_R}
   \notag \\
   & \quad 
   + 
   \frac{ ( \chi' + \mu_R )^2 }{ 4 \lambda_R }
   - 
   \Intk
   \Bigl \{ \,
      \frac{1}{2} \,
      \Bigl [ \,
         \omega_k
         -
         \epsilon_k
         -
         \chi'
         +
         \frac{ |A|^2}{2 \epsilon_k} \,
      \Bigr ]
      \notag \\
      & \quad
      +
      \frac{1}{\beta}
      \Ln{ 1 - e^{-\beta \omega_k} } \,
   \Bigr \} \>.
   \label{J.LOAF.e:pressure-I}
\end{align}

%
%
\subsection{\label{J.s:LOAFphasediag}LOAF phase diagram}

As discussed in our previous papers, there are three possible solutions of the gap equations.  These define three phase-space regions for a given coupling strength.  They are as follows:
\begin{enumerate}
\item[I.] The ``broken symmetry'' case where $\phi \ne 0$.  From \eqref{J.LOAF.e:gapeq-I-a}, if $\phi$ and $\phi^{\ast}$ are nonzero this means that $\chi^{\prime\,2} = |A|^2$.  Because of the $U(1)$ invariance of the Lagrangian, we can choose $\phi$ and $A$ to be real, in which case $\chi' = A$.  In this region $\omega_k = \sqrt{ \epsilon_k \, ( \, \epsilon_k + 2 \chi' \, ) }$.  Here $0 < T < T_c$.

\item[II.] The state when $\phi = 0$ and $0 < | A | < \chi'$.  In this region $\omega_k = \sqrt{ ( \epsilon_k + \chi' )^2 - |A|^2 }$, for $T_c < T < T^\star$. 

\item[III.] The normal state when $\phi = 0$, $|A| = 0$, and $\chi' > 0$, for $T^\star < T$.  In this region $\omega_k = \epsilon_k + \chi'$.
\end{enumerate}

\begin{figure}
   \includegraphics[width=\columnwidth]{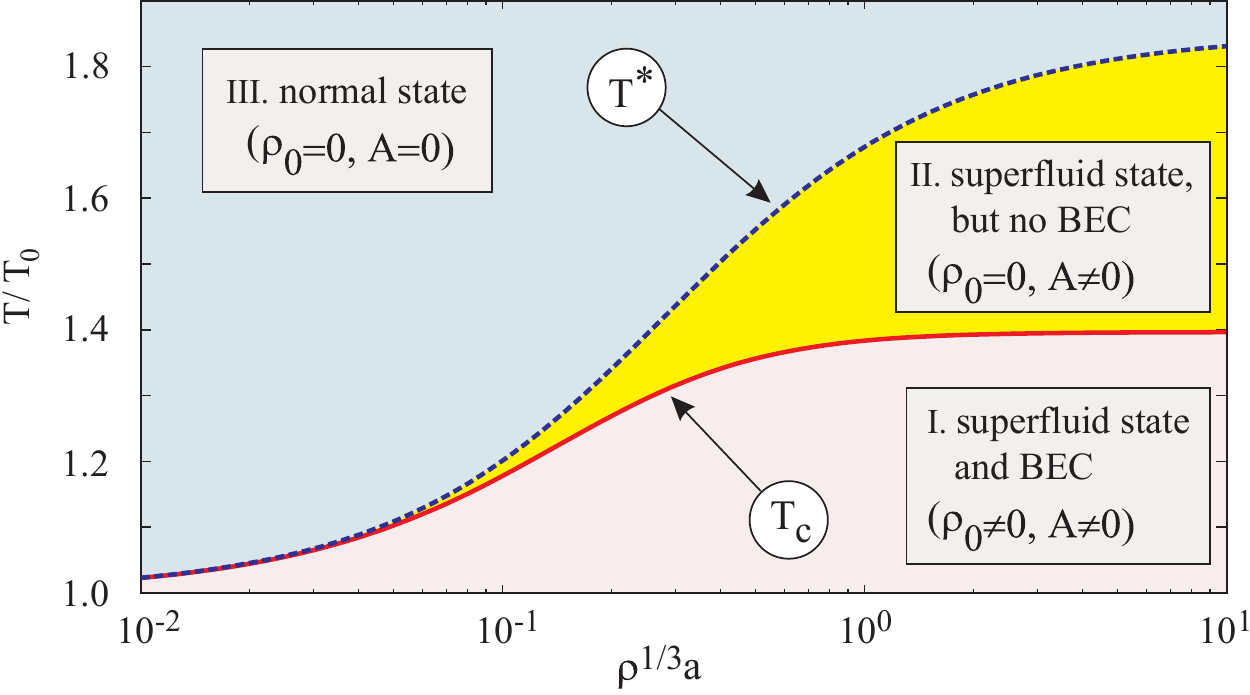}
   \caption{\label{fig:PD}(Color online) 
   LOAF phase diagram.  
   }
\end{figure}

So the LOAF phase diagram shown in Fig.~\ref{fig:PD} is characterized by two special temperatures, $T_c$ and $T^\star$.  The critical temperature $T_c$ corresponds to the appearance of the BEC condensate in the system.  We will show next that the temperature $T^\star$ corresponds to the onset of superfluidity in the system, and is related to the new order parameter, $A$.   

%
%
\section{\label{G.s:goldstone}Goldstone theorem}

The classical Lagrangian density with auxiliary fields and currents from Eq.~\eqref{J.intro.e:S-II} is given by
\begin{align}
   \calL[\Phi,J]
   &=
   \frac{1}{2}
   \Set{ \phi^{\ast}(x), \phi(x) } 
   \begin{pmatrix}
      h^{(+)} & -A(x) \\
      -A^{\ast}(x) & h^{(-)}
   \end{pmatrix}   
   \begin{pmatrix}
      \phi(x) \\ \phi^{\ast}(x)
   \end{pmatrix}
   \notag \\
   & \qquad
   +
   \frac{| A(x) |^2 - \chi^2(x)}{2 \lambda}
   -
   J_{\alpha}(x) \, \Phi^{\alpha}(x) \>,
   \label{J.G.e:lagden}
\end{align}
where $h^{\pm}$ is given in Eq.~\eqref{J.intro.e:hplusminus}.  With the exception of the currents, this Lagrangian is invariant under a global $U(1)$ transformation of the form,
\begin{equation}\label{J.G.e:U1form}
   \phi(x)
   \rightarrow
   e^{i\theta} \, \phi(x) \>,
   \Quad{and}
   A(x)
   \rightarrow
   e^{2i\theta} \, A(x) \>, 
\end{equation}
with $\chi(x)$ unchanged.  
%
%
Consider the change in the Lagrangian density under the infinitesimal change,
\begin{align}
   \Phi^\alpha(x)  
   &\rightarrow 
   \Phi^\alpha(x) + \delta  \Phi^\alpha(x) \>,
   \label{J.G.e:deltaPhi} \\
    \delta  \Phi^\alpha(x) 
    &= 
    \frac{i}{\hbar} \ \epsilon \  g^\alpha{}_\beta \, \Phi^\beta(x) \>.
    \notag
\end{align}
From
\begin{equation*} 
   \delta \calL
   = 
   \frac{\delta \calL}{\delta \Phi_\alpha } \delta \Phi_\alpha 
   + 
   \frac{\delta \calL}{\delta \, \partial_\mu  \Phi_\alpha} \, \delta \, \partial_\mu \Phi_\alpha  
   = 
   \partial_\mu 
   \Bigl ( 
      \frac{\delta \calL}{\delta \, \partial_\mu  \Phi_\alpha}  \, \delta \Phi_\alpha 
   \Bigr ) \>,
\end{equation*}
and using Lagrange's equation, we obtain
\begin{equation}\label{J.G.e:curcons}
   \frac{i}{\hbar} \, \partial_{\tau} \rho(x)
   +
   \bnabla \cdot \bj(x)
   =
   \frac{i}{\hbar} \, \Phi_{\alpha}(x) \, g^{\alpha}{}_{\beta} \, J^{\beta}(x) \>,
\end{equation}
where
\begin{align}
   \rho(x) 
   &= 
   | \phi(x) |^2 \>,
   \label{e:rho0def} \\
   \bj(x) 
   &= 
   \frac{\hbar}{2 i m} \, 
   \bigl [ \, 
      \phi^{\ast}(x) \bnabla \phi(x) 
      -  
      \phi(x) \, \bnabla \phi^{\ast}(x) \, 
   \bigr ] \>.
   \notag
\end{align}
Eq.~\eqref{J.G.e:curcons} is a classical result and represents the $U(1)$ charge conservation equation. Here, we introduced the $U(1)$ charge metric $g^{\alpha}{}_{\beta}$ as the diagonal matrix given by
\begin{equation}\label{J.G.e:gchargemetric}
   g^{\alpha}{}_{\beta}
   =
   \Diag{ 1,-1,0,2,-2} \>.
\end{equation}

%
%

We multiply \eqref{J.G.e:curcons} by $\Exp{-S[\Phi,J]}$, divide by $Z$, and integrate over the fields $\Phi$ to derive a functional equation:
\begin{align}
   &\frac{i}{\hbar} 
   \frac{1}{Z[J]} 
   \int \rD \Phi \, e^{-S[\Phi,J]} \,
   \Phi_{\alpha}(x) \, g^{\alpha}{}_{\beta} \, J^{\beta}(x)
   \label{J.G.e:fuc1eq} \\
   &=
   \frac{1}{Z[J]} 
   \int \rD \Phi \, E^{-S[\Phi,J]} \,
   \Bigl \{ \,
   \frac{i}{\hbar} \, \partial_{\tau} \rho(x)
   +
   \bnabla \cdot \bj(x)
   \Bigr \} \>. 
   \notag  
\end{align}
Integrating \eqref{J.G.e:fuc1eq} over $[ \rd x ]$ and discarding the surface terms gives
\begin{align}
   &\int [\rd x ] \, J^{\beta}(x)  \, g^{\alpha}{}_{\beta}
   \frac{1}{Z[J]} 
   \int \rD \Phi \, 
   \Phi_{\alpha}(x) \, 
   e^{-S[\Phi,J]} 
   \label{J.G.e:func2eq} \\
   & \qquad
   =
   \int [\rd x ] \, 
   J_{\alpha}(x)  \, g^{\alpha}{}_{\beta} \, \Phi^{\beta}[J](x)
   =   
   0 \>.
   \notag
\end{align}
Changing functional variables from $J$ to $\Phi$ and using Eq.~\eqref{J.ea.e:dGammadphidchi}, we can write Eq.~\eqref{J.G.e:func2eq} as
\begin{equation}\label{J.G.e:func3eq}
   \int [\rd x' ] \, 
   \frac{\delta \Gamma[\Phi]}{\delta \Phi^{\beta}(x')} \, g^{\beta}{}_{\gamma} \, \Phi^{\gamma}(x')
   =
   0 \>.
\end{equation}
Differentiating \eqref{J.G.e:func3eq} with respect to $\Phi^{\alpha}(x)$ gives
\begin{align}
   &\int [\rd x' ] \, 
   \frac{\delta^2 \Gamma[\Phi]}{\delta \Phi^{\alpha}(x) \, \delta \Phi^{\beta}(x')} \, g^{\beta}{}_{\gamma} \, \Phi^{\gamma}(x')
   \label{J.G.e:func4eq} \\
   & \qquad
   =
   \int [\rd x' ] \, \calG^{-1}_{\alpha\beta}[\Phi](x,x') \, g^{\beta}{}_{\gamma} \, \Phi^{\gamma}(x')
   =
   -  J_{\beta}(x) \, g^{\beta}{}_{\alpha} \>.
   \notag
\end{align}
Expanding $\calG^{-1,}{}^{\alpha}{}_{\beta}(x,x')$ in a Fourier series,
\begin{align}
   &\calG^{-1,}{}^{\alpha}{}_{\beta}(x,x')
   \label{J.G.e:Gexpand} \\ 
   & \quad
   =
   \frac{1}{\beta} \!
   \Intq \sum_{n=-\infty}^{+\infty}
   \tilde{\calG}^{-1,}{}^{\alpha}{}_{\beta}(\bq,n) \, e^{i [ \bq \cdot ( \bx - \bx') - \omega_n ( \tau - \tau') ] } \>,
   \notag
\end{align}
and for constant fields and in the absence of sources, Eq.~\eqref{J.G.e:func4eq} gives \cite{r:WeinbergI},
\begin{equation}\label{J.G.e:eigeneq}
   \tilde{\calG}^{-1}{}^{\alpha}{}_{\beta}(0,0) \, \bar{\Phi}^{\beta}
   =
   0 \>,
\end{equation}
where 
\begin{align}
   \bar{\Phi}^{\alpha}
   &=
   g^{\alpha}{}_{\mu} \, \Phi^{\mu}
   =
   \Set{ \phi, -\phi^{\ast}, 0, 2 A, - 2 A^{\ast} } \>,
   \label{J.G.e:barPhidefs} \\
   \bar{\Phi}_{\alpha}
   &=
   g_{\alpha}{}^{\mu} \, \Phi_{\mu}
   =
   \Set{ -\phi^{\ast}, \phi, 0, - 2 A^{\ast}, 2 A } \>.
   \notag
\end{align}
For the LOAF approximation, $\tilde{\calG}^{-1,}{}^{\alpha}{}_{\beta}(\bq,0)$ is given by
\begin{equation}\label{J.G.e:calGinv-I}
   \tilde{\calG}^{-1}{}^{\alpha}{}_{\beta}(\bq,0) 
   =
   \begin{pmatrix}
      \epsilon_q + \chi' & -A & \phi & -\phi^{\ast} & 0 \\
      -A^{\ast} & \epsilon_q + \chi' & \phi^{\ast} & 0 & -\phi \\
      \phi^{\ast} & \phi & \gamma(q) & \delta(q) & \delta^{\ast}(q) \\
      -\phi & 0 & \delta^{\ast}(q) & \alpha(q) & \beta(q) \\
      0 & -\phi^{\ast} & \delta(q) & \beta^{\ast}(q) & \alpha(q)
   \end{pmatrix} \>,
\end{equation}
where
\begin{align}\label{J.G.e:alphabetadefs}
   \alpha(q)
   &=
   \tilde{D}^{-1,1}{}_{1}[\chi_0](\bq,0)
   =
   \tilde{D}^{-1,2}{}_{2}[\chi_0](\bq,0) \>,
   \\
   \beta(q)
   &=
   \tilde{D}^{-1,1}{}_{2}[\chi_0](\bq,0)
   =
   \tilde{D}^{-1,2}{}_{1}[\chi_0]^{\ast}(\bq,0) \>,
   \notag \\
   \gamma(q)
   &=
   \tilde{D}^{-1,0}{}_{0}[\chi_0](\bq,0) \>,
   \notag \\
   \delta(q)
   &=
   \tilde{D}^{-1,0}{}_{1}[\chi_0](\bq,0)
   =
   \tilde{D}^{-1,0}{}_{2}[\chi_0]^{\ast}(\bq,0)
   \notag \\
   &=
   \tilde{D}^{-1,1}{}_{0}[\chi_0]^{\ast}(\bq,0)
   =
   \tilde{D}^{-1,2}{}_{0}[\chi_0](\bq,0) \>.
\end{align}
For small momentum values, the momentum expansion of the Green functions is
\begin{align}
   \alpha(q) &= \alpha_0 + \alpha_1 q^2 + \dotsb \>,
   \\
   \beta(q) &= \beta_0 + \beta_1 q^2 + \dotsb \>,
   \notag \\
   \gamma(q) &= \gamma_0 + \gamma_1 q^2 + \dotsb \>,
   \notag \\
   \delta(q) &= \delta_0 + \delta_1 q^2 + \dotsb \>.
   \notag
\end{align}
Since $\bar{\Phi}^{\beta}$ is missing the term proportional to $\chi$, we can extract the $4 \times 4$ matrix $M^{\alpha}{}_{\beta}$, which is given by
\begin{align}
   M^{\alpha}{}_{\beta}(\bq,0)
   &=
   M_0^{\alpha}{}_{\beta} + M_1^{\alpha}{}_{\beta} \, q^2 + \dotsb
   \label{J.G.e:MLOAF-I} \\
   &=
   \begin{pmatrix}
      \epsilon_q + \chi' & -A & -\phi^{\ast} & 0 \\
      -A^{\ast} & \epsilon_q + \chi' & 0 & -\phi \\
      -\phi & 0 & \alpha(q) & \beta(q) \\
      0 & -\phi^{\ast} & \beta^{\ast}(q) & \alpha(q)
   \end{pmatrix}
   \notag 
\end{align}
where
\begin{equation}\label{J.m2.e:M0derv}
   M_0^{\alpha}{}_{\beta}(0,0)
   =
   \frac{ \partial^2 \, V_{\text{eff}}[\Phi] }{ \partial \Phi_{\alpha} \, \partial \Phi^{\beta} } \>,
\end{equation}
From Eq.~\eqref{J.LOAF.e:Gamma-IV}, we find
\begin{subequations}\label{J.G.e:alphabetadefs-0}
\begin{align}
   \alpha_0
   &=
   \Bigl \{ \,
      \frac{2 n(\beta\omega_k) + 1}{2\omega_k}
      -
      \frac{1}{2\epsilon_k} \,
   \Bigr \}
   \label{J.G.e:alphadef} \\
   & \quad
   +
   \frac{|A|^2}{2} \Intk \, 
   \frac{1}{4 \omega_k} \,
   \Bigl \{ \,
      2 \omega_k \, n(\beta\omega_k) \, (  n(\beta\omega_k) + 1 ) 
      \notag \\
      & \qquad\qquad\qquad
      + 
      2 n(\beta\omega_k) + 1 \,
   \Bigr \} \>,
   \notag \\
   \beta_0
   &=
   \frac{A^2}{2} \Intk \, 
   \frac{1}{4 \omega_k} \,
   \Bigl \{ \,
      2 \omega_k \, n(\beta\omega_k) \, (  n(\beta\omega_k) + 1 ) 
      \notag \\
      & \qquad\qquad\qquad
      + 
      2 n(\beta\omega_k) + 1 \,
   \Bigr \} \>.
   \label{J.G.e:betadef} 
\end{align}
\end{subequations}
Then the determinant of $M_0$ is 
\begin{align}
   \Det{ M_0 }
   &=
   ( \, \alpha_0^2 - |\beta_0|^2 \, ) \, ( \, \chi^{\prime\,2} - |A|^2 \, )
   -
   2 \, \alpha_0 \, \chi' \, |\phi|^2
   \notag \\
   & \qquad
   +
   \beta_0^{\ast} \, A \, \phi^2 + \beta_0 \, A^{\ast} \, \phi^{\ast\,2}
   +
   |\phi|^4 \>.
   \label{J.G.e:detGresult} 
\end{align}
From the gap Eq.~\eqref{J.LOAF.e:gapeq-I-c} and for real fields we find that
\begin{equation}\label{J.G.e:xxx1}
   \alpha_0 - \beta_0
   =
   \frac{ \phi^2 }{ 2 A } \>.
\end{equation}
The Goldstone theorem corresponds to $\Det{M_0} = 0$.  Hence, LOAF shows that Goldstone modes are present in region I and II of the phase diagram.  The condition $\Det{M_0} = 0$ is satisfied in region I because $\chi' = |A|$, and $\phi \ne 0$, whereas in region II we have $\phi=0$ and $A \ne 0$ and the condition $\Det{M_0} = 0$ is satisfied also.  It can be shown that the determinant of the full $5 \times 5$ inverse Green function also vanishes at $\bq = 0$ and $n=0$, $\Det{ \tilde{\calG}^{-1}{}^{\alpha}{}_{\beta}(0,0) } = 0$, when $\Det{M_0} = 0$.

%
%
\section{\label{s:landaucalc}Superfluid density in a Bose gas of particles}

Traditionally, the density of the superfluid component of the Bose gas is calculated using Landau's phenomenological two-fluid theory, as shown for instance in Ref.~\onlinecite{r:Fetter:1971fk}.  In this theory, the superfluid density is given by
\begin{equation}\label{J.e:rhos}
   \rho_s
   =
   \rho - \rho_n \>,
\end{equation}
where $\rho$ is the total density, $\rho_s$ is the superfluid density, and the normal density, $\rho_n$, defined as
\begin{equation}\label{J.e:rhon}
   \rho_n
   =
   \frac{\hbar^2}{6\pi^2}
   \int_{0}^{\infty}
   \!\!\! \rd k \, k^4 \,
   \Bigl [ \,
      - \frac{\partial n(\beta\omega_k)}{\partial \omega_k} \,
   \Bigr ] 
   \>,
\end{equation}
where $n(x) = [ \, 1 - e^{-x} \, ]^{-1}$ is the particle number density, and the dispersion relation is given by
\begin{equation}\label{J.e:omegadef}
   \omega_k^2
   =
   ( \, \epsilon_k + \chi' \, )^2 - A^2 \>,
   \qquad
   \epsilon_k
   =
   \frac{\hbar^2 k^2}{2m} \>.
\end{equation}
In this section, $\beta = 1/(\kBolt T)$.
We will show next that the superfluid density is proportional to the square of the order parameter $A$.  To this end, we formally consider the series expansion of the normal density in powers of the order parameter,
\begin{equation}\label{J.e:rhon-exp}
   \rho_n
   =
   \rho - \kappa \, A^2 + \dotsb \>,
\end{equation}
so that
\begin{equation}\label{PT.e:rhos}
   \rho_s
   =
   \kappa \, A^2 + \dotsb
\end{equation}
We begin by integrating Eq.~\eqref{J.e:rhon} by parts. We use
\begin{equation}\label{J.e:identityI}
   \frac{\partial n(\beta\omega_k)}{\partial \omega_k}
   =
   \frac{\omega_k}{\epsilon_k + \chi} \, \frac{m}{\hbar^2 k} \, \frac{\partial n(\beta\omega_k)}{\partial k}
   \rightarrow
   \frac{m}{\hbar^2 k} \, \frac{\partial n(\beta\omega_k)}{\partial k} \>,
\end{equation}
when $A \rightarrow 0$, to write Eq.~\eqref{J.e:rhon} as
\begin{align}
   \rho_n
   \label{J.a.e:rhon-I}
   =
   - \frac{m}{6\pi^2}
   \int_{0}^{\infty}
   \!\!\! \rd k \, 
   \frac{ k^3 \, \omega_k }{ \epsilon_k + \chi' } \,
   \frac{\partial n(\beta\omega_k)}{\partial k} \>.
\end{align}
Integrating by parts gives
\begin{align}
   &\rho_n
   =
   \frac{1}{6\pi^2}
   \Bigl ( \frac{m}{\hbar} \Bigr )^{\!2}
   \label{J.a.e:rhon-II} \\
   & \qquad
   \times
   \int_{0}^{\infty}
   \!\!\! \rd k \, n(\beta\omega_k) \,
   \Bigl \{ \,
      \frac{ 6 \, \epsilon_k \, \omega_k }{ \epsilon_k + \chi' }
      +
      \frac{ 4 \, \epsilon_k^2 \, | A |^2 }{ \omega_k \, ( \epsilon_k + \chi' )^2 } \,
   \Bigr \} \>.
   \notag
\end{align}
The second term in the last line of \eqref{J.a.e:rhon-II} is already of order $A^2$, and in the limit $A \rightarrow 0$ gives a term
\begin{equation}\label{J.a.e:secondterm}
   \frac{1}{(2\pi)^2}
   \Bigl ( \frac{m}{\hbar} \Bigr )^{\!2}
   \int_{0}^{\infty}
   \!\!\! \rd k \, \frac{n(\beta\omega_{0\,k})}{\epsilon_k + \chi} \,
   \Bigl ( \frac{8}{3} \Bigr )  \,
   \frac{ \epsilon_k^2 }{ (\epsilon_k + \chi')^2 } \, | A |^2 \>.
\end{equation}
For the first term, since $n_k$ involves a term of order $A^2$, we need to expand it in a power series in $A^2$:
\begin{equation}\label{J.a.e:nkexpand}
   n(\beta\omega_k)
   =
   n(\beta\omega_{0\,k})
   +
   \Bigl [ \frac{\partial n(\beta\omega_k)}{\partial A^2} \Bigr ]_{A^2=0} \, A^2
   +
   \dotsb \>.
\end{equation}
We use
\begin{equation}\label{J.a.e:use}
   \rd (A^2)
   =
   -
   \rd (\omega_k^2 )
   =
   - 2 \, ( \, \epsilon_k + \chi' \, ) \, \Bigl ( \frac{\hbar^2 k}{m} \bigr ) \, \rd k \>,
\end{equation}
so that
\begin{equation}\label{J.a.e:partialA2partialk}
   \frac{\partial n(\beta\omega_k)}{\partial A^2}
   =
   -
   \frac{1}{2} \, \Bigl ( \frac{m}{\hbar^2} \bigr ) \, \frac{1}{ k \, ( \epsilon_k + \chi' \, ) } \, \frac{\partial n(\beta\omega_k)}{\partial k} \>.
\end{equation}
this gives then for the \emph{first} term,
\begin{align}
   &\frac{1}{\pi^2}
   \Bigl ( \frac{m}{\hbar} \Bigr )^{\!2}
   \int_{0}^{\infty}
   \!\!\! \rd k \, 
   \frac{\epsilon_k \, \omega_k }{ \epsilon_k + \chi' } \,
   \Bigl \{ \,
      n(\beta\omega_{0\,k})
      \label{J.a.e:firstI} \\
      & \qquad
      -
      \frac{1}{2} \, \Bigl ( \frac{m}{\hbar^2} \bigr ) \, \frac{A^2}{ k \, ( \epsilon_k + \chi' \, ) } \, \frac{\partial n(\beta\omega_k)}{\partial k}
      +
      \dotsb \,
   \Bigr \}
   \notag \\
   &=
   \rho
   - 
   \kappa' \, A^2
   +
   \dotsb \>,
   \notag
\end{align}
where $\kappa'$ is given by
\begin{align}
   \kappa'
   &=
   \frac{1}{(2\pi)^2}
   \Bigl ( \frac{m}{\hbar} \Bigr )^{\!2}
   \int_{0}^{\infty}
   \!\!\! \rd k \, 
   \frac{ k \, \omega_k }{ ( \epsilon_k + \chi' )^2 } \,
   \frac{\partial n_k}{\partial k}
   \label{J.a.e:alphap} \\
   &=
   - \frac{1}{(2\pi)^2}
   \Bigl ( \frac{m}{\hbar} \Bigr )^{\!2}
   \int_{0}^{\infty} \!\!\! \rd k \,
   \frac{ n(\beta\omega_{0\,k}) }{ \epsilon_k + \chi' } \,
   \Bigl [ \,
      1
      -
      \frac{2 \, \epsilon_k}{ \epsilon_k + \chi' } \,
   \Bigr ] \>,
   \notag
\end{align}
where in the last line we have integrated by parts.  Adding this result to what we found in Eq.~\eqref{J.a.e:secondterm} gives
\begin{align}
   \kappa
   &=
   -
   \frac{\hbar^2}{(2\pi)^2} \,
   \int_{0}^{\infty}
   \!\!\! \rd k \, 
   \frac{n(\beta\omega_k)}{\epsilon_k + \chi'}
   \label{J.a.e:alpha-final} \\
   & \qquad
   \times
   \Bigl [ \,
      1
      -
      \frac{2 \, \epsilon_k}{\epsilon_k + \chi'}
      +
      \frac{8}{3} \,
      \frac{\epsilon_k^2}{( \epsilon_k + \chi' )^2} \,
   \Bigr ] \>.
   \notag
\end{align}
This completes our derivation.

%
%
\section{\label{s:microcalc}Microscopic theory of the superfluid density}

Consider a superfluid moving with velocity $\bv$ in the laboratory frame.  The Lagrangian for this system is obtained by replacing the momentum operator for the system at rest by
\begin{equation}\label{J.m1.e:momop}
   \frac{\hbar}{i} \, \bnabla
   \mapsto
   \bP
   \equiv
   \frac{\hbar}{i} \, \bnabla
   -
   m \, \bv \>.
\end{equation}
Then, the superfluid mass density is given by the second-order derivative of the free energy,
\begin{equation}\label{J.m.e:rhosFreeEnergy}
   \rho_s
   =
   \frac{1}{V} \, 
   \Bigl [ \,
      \frac{ \partial^2 \, F(V,N,T,v) }{ \partial v^2 } \,
   \Bigr ]_{v=0} \>,
\end{equation}
where $v$ is the velocity of the superfluid.  The free energy is related to the grand potential $\Omega(V,\mu,T,v)$ by
\begin{equation}\label{J.m.e:FtoOmega}
   F(V,N,T,v) = \Omega(V,\mu,T,v) + \mu \, N \>.
\end{equation}
It was shown in Ref.~\onlinecite{PhysRevA.74.063626} that Eq.~\eqref{J.m.e:rhosFreeEnergy} is equivalent to
\begin{equation}\label{J.m.e:rhos}
   \rho_s
   =
   \frac{1}{V} \, 
   \Bigl [ \,
      \frac{ \partial^2 \, \Omega(v) }{ \partial v^2 } \,
   \Bigr ]_{v=0} \>,
\end{equation}
which is what we use here.

With respect to the laboratory frame, the Euclidian Lagrangian is given by
\begin{equation}\label{J.m1.e:calLdef}
   \calL
   =
   \frac{1}{2} \,
   \Bigl [ \,
      \phi^{\ast}(x) \, 
      \frac{\partial \phi(x)}{\partial \tau}
       -
      \phi(x) \,
      \frac{\partial \phi^{\ast}(x)}{\partial \tau} \, 
   \Bigr ]
   +
   \calH \>,
\end{equation}
where $\calH$ is given by
\begin{align}
   \calH
   &=
   \phi^{\ast}(x) \, 
   \Bigl [ \,
      -
      \frac{ \hbar^2 \nabla^2}{2m}
      -
      \frac{\hbar}{i} \, \bv \cdot \bnabla
      +
      \frac{1}{2} \, m \, v^2 
      -
      \mu \,
   \Bigr ] \, 
   \phi(x)
   \notag \\
   & \qquad\qquad
   +
   \frac{\lambda}{2} \,
   | \, \phi(x) \, |^4 \>,
   \label{J.m1.e:calHdef}
\end{align}   
where the fields $\phi(x)$ are given in Lab coordinates.  Introducing the Hubbard-Stratonovitch transformation by adding the auxiliary Lagrangian
\begin{align}
   \calL_{\text{aux}}
   &=
   -
   \frac{1}{2 \lambda} \,
   \bigl [ \,
      \chi(x) - \lambda \, \sqrt{2} \, | \phi(x) |^2 \,
   \bigr ]^2
   \label{J.m1.e:Laux} \\
   & \quad
   +
   \frac{1}{2 \lambda} \,
   \bigl | \,
      A(x) 
      - 
      \lambda \, \phi^{2}(x) \,
   \bigr |^2 \>,
   \notag
\end{align}
to $\calL$ gives
\begin{align}
   \calL'
   &=
   \calL + \calL_{\text{aux}}
   \label{J.m1.e:calLvp} \\
   &=
   \frac{1}{2} \,
   \Bigl [ \,
      \phi^{\ast}(x) \, 
      \frac{\partial \phi(x)}{\partial \tau}
       -
      \phi(x) \,
      \frac{\partial \phi^{\ast}(x)}{\partial \tau} \, 
   \Bigr ]
   +
   \calH' \>,
\end{align}
where now
\begin{align}
   \calH'
   &=
   \phi^{\ast}(x) \, 
   \Bigl [ \,
      -
      \frac{ \hbar^2 \nabla^2}{2m}
      -
      \frac{\hbar}{i} \, \bv \cdot \bnabla
      +
      \frac{1}{2} \, m \, v^2 
      \label{J.m1.e:Hvp} \\
      & \qquad\qquad\qquad
      +
      \sqrt{2} \, \chi(x)
      -
      \mu \,
   \Bigr ] \, 
   \phi(x)
   \notag \\
   & \qquad
   -
   \frac{1}{2} \,
   \bigl [ \,
      A(x) \, [ \phi^{\ast}(x) ]^2
      +
      A^{\ast}(x) \, [ \phi(x) ]^2 \,
   \bigr ] 
   \notag \\
   & \qquad
   +
   \frac{1}{2\lambda} \,
   \bigl [ \,
      | A(x) |^2 - \chi^2(x) \,
   \bigr ] \>.
   \notag
\end{align}
Adding currents, the action becomes
\begin{align}
   &S[J,\Phi;\mu,\beta,\bv]
   \label{J.m1.e:Sv-II} \\
   & \quad
   =
   \frac{1}{2}
   \iint [\rd x] \, [\rd x'] \,
   \phi_{a}(x) \, G^{-1}_v{}^a{}_b[\Phi](x,x') \, \phi^{b}(x')
   \notag \\
   & \quad
   +
   \int [\rd x] \,
   \bigl \{ \,
      [ \, | A(x) |^2 - \chi^2(x) \, ] / (2 \lambda )
      +
      J^{\alpha}(x) \, \Phi^{\alpha}(x) \,
   \bigr \} \>,
   \notag
\end{align}
where 
\begin{equation}\label{J.m1.e:Gvinvdef}
   G^{-1}_v{}^a{}_b[\Phi](x,x')
   =
   \delta(x,x') \,
   \begin{pmatrix}
      h_v^{(+)} & -A(x) \\
      -A^{\ast}(x) & h_v^{(-)}
   \end{pmatrix} \>,
\end{equation}
with
\begin{gather}
   h_v^{(+)}
   =
   h_v + \partial_{\tau} \>,
   \qquad
   h_v^{(-)}
   =
   h_v^{\ast} - \partial_{\tau} \>,
   \label{J.m1.e:hplusminus} \\
   h_v
   =
   -
   \frac{ \hbar^2 \nabla^2}{2m}
   -
   \frac{\hbar}{i} \, \bv \cdot \bnabla
   +
   \frac{1}{2} \, m \, v^2
   +
   \sqrt{2} \, \chi(x)
   -
   \mu \>.
   \notag    
\end{gather}
Note that in the laboratory frame, the dependence of the action on the superfluid velocity is contained entirely in the inverse Green function $G^{-1}_v{}^a{}_b[\Phi](x,x')$.  So computing derivative of the action, we find
\begin{equation}\label{J.m1.e:dSdvi_II}
   \frac{\partial S}{\partial v_i}
   =
   - \int [\rd x] \, j_i(x) \>,
\end{equation}
where $j_i(x)$ is the classical mass current density,
\begin{align}
   j_i(x)
   &=
   \frac{\hbar}{2i} \,
   \bigl [ \,
      \phi^{\ast}(x) \, \nabla_i \, \phi(x)
      -
      \phi(x) \, \nabla_i \, \phi^{\ast}(x) \,
   \bigr ]
   -
   v_i \, \rho(x) \>,
   \label{J.m1.e:jxclass}
\end{align}
with $\rho(x) = | \phi(x) |^2$.
The second derivative of $S$ with respect to the superfluid velocity is simply
\begin{equation}\label{J.m1.e:d2Sdvdv}
   \frac{\partial^2 S}{\partial v_i \, \partial v_j}
   =
   \delta_{ij} \int [\rd x] \, \rho(x) \>.
\end{equation}
The partition function $Z[J;V,\mu,\beta,\bv]$ is also a function of the velocity $\bv$.
Then
\begin{align}
   \frac{1}{Z} \frac{\partial Z}{\partial v_i}
   &=
   - \frac{\calN}{Z}
   \int \rD \Phi \,
   \frac{\partial S}{\partial v_i} \, e^{-S}
   \label{J.m1.e:dZdj} \\
   &=
   \int [\rd x] \, \Expect{ j_i(x) } \>.
   \notag
\end{align}
and
\begin{align}
   &\frac{1}{Z} \frac{\partial^2 Z}{\partial v_i \, \partial v_j}
   =
   \frac{\calN}{Z}
   \int \rD \Phi \,
   \Bigl [ \,
      \Bigl ( \frac{\partial S}{\partial v_i} \Bigr ) \,
      \Bigl ( \frac{\partial S}{\partial v_j} \Bigr )
      -
      \frac{\partial^2 S}{\partial v_i \, \partial v_j} \,
   \Bigr ] \, e^{-S}
   \notag \\
   & \quad
   =
   \frac{\calN}{Z}
   \iint [\rd x] \, [\rd x'] 
   \int \rD \Phi
   \label{J.m1.e:dZdjdj} \\
   & \qquad\qquad
   \times
   \Bigl [ \,
      j_i(x) \, j_j(x')
      -
      \delta_{ij} \delta(x,x') \, \rho(x) \,
   \Bigr ] \, e^{-S}
   \notag \\
   &=
   \iint [\rd x] \, [\rd x'] 
   \Expect{ j_i(x) \, j_j(x') - \delta_{ij} \delta(x,x') \, \rho(x) } \>,
   \notag
\end{align}
However, we need derivatives with respect to the grand potential.  These are given by
\begin{align}
   \frac{\partial Z}{\partial v_i}
   &=
   - 
   \beta Z \, \frac{\partial \Omega}{\partial v_i} \>,
   \label{J.m1.e:dZs} \\
   \frac{\partial^2 Z}{\partial v_i \, \partial v_j}
   &=
   - 
   \beta Z \, \frac{\partial^2 \Omega}{\partial v_i \, \partial v_j}
   +
   \beta^2 Z \, \frac{\partial \Omega}{\partial v_i} \, \frac{\partial \Omega}{\partial v_j}  \>.
   \notag
\end{align}
So we find
\begin{equation}\label{J.m1.e:dOmegas}
   \frac{\partial \Omega}{\partial v_i} 
   =
   - \frac{1}{\beta Z} \, \frac{\partial Z}{\partial v_i} 
   =
   - \frac{1}{\beta} \, \int [\rd x] \, \Expect{ j_i(x) } \>,
\end{equation}
and
\begin{align}
   &\frac{\partial^2 \Omega}{\partial v_i \, \partial v_j}
   =
   \beta \, \frac{\partial \Omega}{\partial v_i} \, \frac{\partial \Omega}{\partial v_j}
   -
   \frac{1}{\beta Z} \, \frac{\partial^2 Z}{\partial v_i \, \partial v_j}
   \label{J.m1.e:ddOmegas} \\
   & \quad
   =
   \frac{1}{\beta Z^2} \, \frac{\partial Z}{\partial v_i} \, \frac{\partial Z}{\partial v_j}
   -
   \frac{1}{\beta Z} \, \frac{\partial^2 Z}{\partial v_i \, \partial v_j} \>.
   \notag \\
   & \quad
   =
   \frac{1}{\beta} \delta_{ij}
   \int [\rd x] \, \Expect{\rho(x)}
   \notag \\
   & 
   -
   \frac{1}{\beta} 
   \iint [\rd x] \, [\rd x'] \,
   \bigl [ \,
      \Expect{ j_i(x) \, j_j(x') }
      -
      \Expect{j_i(x)} \, \Expect{j_j(x)} \,
   \bigr ] \>.
   \notag
\end{align}
Assuming for simplicity that the superfluid is moving in the $z$~direction, we obtain the superfluid density from Eq.~\eqref{J.m.e:rhos}, by evaluating Eq.~\eqref{J.m1.e:ddOmegas} for $i=j=z$. We obtain
\begin{equation}\label{J.m1.e:evalzz}
   \rho_s
   =
   \frac{1}{V}
   \Bigl [ \,
      \frac{ \partial^2 \, \Omega(v) }{ \partial v_z^2 } \,
   \Bigr ]_{v=0}
   =
   \rho - \rho_n \>,
\end{equation}
where $\rho$ is the total mass density given by
\begin{equation}\label{J.m1.e:rhodensity}
   \rho
   =
   \frac{1}{\beta V} \int [ \rd x ] \, \Expect{\rho(x)} \>,
\end{equation}
and $\rho_n$ is the normal mass density,
\begin{equation}\label{J.m1.e:rhonjj}
   \rho_n
   =
   \frac{1}{\beta V} 
   \iint [\rd x] \, [ \rd x' ] \, j_{zz}(x,x') \>,
\end{equation}
where now 
\begin{equation}\label{J.m1.e:defijcurrent}
   j_{ij}(x,x')
   =
   \Expect{ j_i(x) \, j_j(x') } - \Expect{j_i(x)} \Expect{j_j(x')}
\end{equation}
is to be evaluated at $v=0$.  That is, the action $S$ now is evaluated at $v=0$ and the current is given by
\begin{equation}\label{J.m1.e:curvzero}
   j_i(x)
   =
   \frac{\hbar}{2 i} \,
   \bigl [ \,
      \phi^{\ast}(x) \, \nabla_i \, \phi(x)
      -
      \phi(x) \, \nabla_i \, \phi^{\ast}(x) \,
   \bigr ] \>,
\end{equation}
without the additional $v_i \, \rho(x)$ term.

The calculation of the normal density from \eqref{J.m1.e:rhonjj} requires the evaluation of the four-point correlation function,
\begin{align}
   &\Expect{ T_{\tau}\{ \, \hat{\phi}^{a}(x_1)\, \hat{\phi}^{b}(x_2)\, \hat{\phi}^{c}(x_3)\, \hat{\phi}^{d}(x_4) \} }
   \\ &\qquad\qquad
   =
   \frac{1}{Z} \, \frac{\delta^4 Z[J]}{\delta j_{a}(x_1) \, j_{b}(x_2) \, j_{c}(x_3) \, j_{d}(x_4) } \>,
   \notag 
\end{align}
which in leading order in $\epsilon$ is given by the products of two-point functions,
\begin{align}
   & G^{ab}(x_1,x_2) \, G^{cd}(x_3,x_4) 
   \\ & \notag \quad
   +
   G^{ac}(x_1,x_3) \, G^{bc}(x_2,x_4)
   +
   G^{bc}(x_2,x_3) \, G^{ad}(x_1,x_4) \>.   
\end{align}
In this approximation, the calculation of the current-current correlation function follows a straightforward path.  We obtain 
\begin{align}
   \rho_n
   &=
   \tilde{j}_{zz}(0,0) 
   \label{J.c.e:rhonjj-III} \\
   &=
   \frac{\hbar^2}{6\pi^2} \int_{0}^{\infty} \!\! k^4 \rd k \,
   \Bigl [ \,
      - \frac{ \partial \, n(\beta\omega_k) }{\partial \omega_k} \,
   \Bigr ] \>,
   \notag
\end{align}
where $\tilde{j}_{ij}(\bq,s)$ is the Fourier transform of the current-current correlation function in the sense of
\begin{equation}\label{J.c.e:jexpand}
   j_{ij}(x,y)
   =
   \frac{1}{\beta}
   \Intq \sum_{s=-\infty}^{+\infty}
   \tilde{j}_{ij}(\bq,s) \, 
   e^{i [ \,\bq \cdot (\bx - \by) - \omega_s ( \tau_x - \tau_y) \,]} \>.
\end{equation}
Eq.~\eqref{J.c.e:rhonjj-III} agrees with Landau's classical formula in Eq.~\eqref{J.e:rhon}.

%
%
\section{\label{J.m.s:twist}Josephson relation}

The calculation of the superfluid density discussed in the previous section was performed in the laboratory frame.  The dependence of the action on the superfluid velocity is contained entirely in the inverse Green function, as discussed previously.  This dependence can be removed by changing variables to a new field $\phi_v(x)$, defined by
\begin{equation}\label{J.m2.e:Galtransforms}
   \phi_v(x)
   =
   e^{ i ( m \bv \cdot \br )/\hbar} \, \phi(x) \>, 
\end{equation}
which is sometimes called a ``twist'' transformation \cite{r:Fisher:1973fk}.  
Then since
\begin{equation}\label{J.m2.e:UgradU}
   e^{ - i ( m \bv \cdot \br )/\hbar} \, 
   \Bigl [ \,
      \frac{\hbar}{i} \, \bnabla - m \, \bv \,
   \Bigr ] \,
   e^{ + i ( m \bv \cdot \br )/\hbar}
   =
   \frac{\hbar}{i} \, \bnabla \>, 
\end{equation}
in terms of the fields $\phi_v(x)$, the action \eqref{J.m1.e:Sv-II} in the Lab frame becomes
\begin{align}
   &S[J,\Phi_v;\mu,\beta,\bv]
   \label{J.m2.e:Sv-III} \\
   & \quad
   =
   \frac{1}{2}
   \iint [\rd x] \, [\rd x'] \,
   \phi_{v\,a}(x) \, G^{-1}{}^a{}_b[\Phi_v](x,x') \, \phi^{b}_v(x')
   \notag \\
   & \qquad
   +
   \int [\rd x] \,
   \bigl \{ \,
      [ \, | A_v(x) |^2 - \chi^2(x) \, ] / (2 \lambda )
      +
      J^{\alpha}(x) \, \Phi_v^{\alpha}(x) \,
   \bigr \} \>,
   \notag
\end{align}
where now
\begin{equation}\label{J.m2.e:Gvinvdef-III}
   G^{-1}{}^a{}_b[\Phi_v](x,x')
   =
   \delta(x,x') \,
   \begin{pmatrix}
      h^{(+)} & -A_v(x) \\
      -A_v^{\ast}(x) & h^{(-)}
   \end{pmatrix} \>,
\end{equation}
with
\begin{gather}
   h^{(+)}
   =
   h + \partial_{\tau} \>,
   \qquad
   h^{(-)}
   =
   h - \partial_{\tau} \>,
   \label{J.m2.e:hplusminus-III} \\
   h
   =
   -
   \frac{ \hbar^2 \nabla^2}{2m}
   +
   \sqrt{2} \, \chi(x)
   -
   \mu \>,
   \notag    
\end{gather}
and we have defined the auxiliary field $A_v(x)$ as
\begin{equation}\label{J.m2.e:Acomoving}
   A_v(x)
   =
   e^{ i ( 2 m \bv \cdot \br ) / \hbar } \, A(x) \>.
\end{equation}
Here the new fields and currents depend on the superfluid velocity $\bv$ and are given by
\begin{subequations}\label{K.m2.e:PhiJdefs-I}
\begin{align}
   \Phi_v^{\alpha} 
   &= 
   \Set{\phi_v,\phi_v^{\ast},\chi,A_v,A_v^{\ast}} \>,
   \label{J.m2.e:Phidefs-I} \\
   \Phi_{v\,\alpha} 
   &= 
   \Set{\phi_v^{\ast},\phi_v,\chi,A_v^{\ast},A_v} \>,
   \notag \\
   J_v^{\alpha}(x)
   &=
   \Set{ j_v,j_v^{\ast},s,S_v,S_v^{\ast}} \>,
   \label{J.m2.e:Jdefs-I} \\
   J_{v\,\alpha}(x)
   &=
   \Set{ j_v^{\ast},j_v,s,S_v^{\ast},S_v} \>,
   \notag   
\end{align}
\end{subequations}
with similar definitions for the new currents $J_v(x)$.  The important thing to note here is that the entire dependence on $\bv$ now resides in the fields rather that in the Green function operator.  In particular, we have
\begin{equation}\label{J.m2.e:dPhidv}
   \Bigl ( \frac{\hbar}{i}  \Bigr ) \, \frac{\partial}{\partial v_i} \, \Phi_v^{\alpha}(x)
   =
   m x_i \, g^{\alpha}{}_{\beta} \, \Phi_v^{\beta}(x) \>,
\end{equation}
where
\begin{equation}\label{J.m2.e:chargedef}
   g^{\alpha}{}_{\beta}
   =
   \Set{ 1,-1,0,2,-2 } \>.
\end{equation}
The currents also now depend on $\bv$, but in the following derivation, we will set the currents to zero.
So then
\begin{align}
   \Bigl ( \frac{\hbar}{i}  \Bigr ) \, \frac{\partial S}{\partial v_i}
   &=
   \int [\rd x] \, \Bigl ( \frac{\hbar}{i}  \Bigr ) \, \frac{\partial \Phi_v^{\alpha}(x)}{\partial v_i} \, 
   \frac{ \delta S[\Phi_v]}{\delta \Phi_v^{\alpha}(x) }
   \label{J.m2.e:dSdv} \\
   &=
   \int [\rd x] \,  m x_i \, g^{\alpha}{}_{\mu} \,
   \frac{ \delta S[\Phi_v]}{\delta \Phi_v^{\alpha}(x) } \, \Phi_v^{\mu}(x) \>.
   \notag
\end{align}
and
\begin{align}
   &\Bigl ( \frac{\hbar}{i}  \Bigr )^2 \, \frac{\partial^2 S}{\partial v_i \, \partial v_j}
   =
   m^2 \iint [\rd x] \, [\rd x'] \,
   x_i \, x'_j \,
   \label{J.m2.e:d2Sdvdv} \\
   & \quad
   \times
   \Bigl \{ \,
      \delta(x,x') \, \frac{\delta S[\Phi_v]}{\delta \Phi_v^{\alpha}(x)} \, \Phi_v^{\alpha}(x)
      \notag \\
      & \qquad
      +
      g^{\alpha}{}_{\mu} \, g^{\beta}{}_{\nu} \, 
      \frac{\delta^2 S[\Phi_v]}{\delta \Phi_v^{\alpha}(x) \, \delta \Phi_v^{\beta}(x')} \,
      \Phi_v^{\mu}(x) \,  \Phi_v^{\nu}(x') \,
   \Bigr \} \>.
   \notag
\end{align}
So derivatives of the partition function are now given by
\begin{align}
   &\Bigl ( \frac{\hbar}{i}  \Bigr ) \, \frac{1}{Z} \frac{\partial Z}{\partial v_i}
   =
   - \frac{\calN}{Z}
   \int \rD \Phi \,
   \Bigl ( \frac{\hbar}{i}  \Bigr ) \,
   \frac{\partial S}{\partial v_i} \, e^{-S}
   \label{J.m2.e:dZdj} \\
   & \quad
   =
   - m \int [\rd x] \, x_i \,  g^{\alpha}{}_{\mu} \, 
   \Expectbig{ \frac{ \delta S[\Phi_v]}{\delta \Phi_v^{\alpha}(x) } \, \Phi_v^{\mu}(x) } \>,
   \notag
\end{align}
and
\begin{align}
   &\Bigl ( \frac{\hbar}{i}  \Bigr )^2 \, \frac{1}{Z} \frac{\partial^2 Z}{\partial v_i \, \partial v_j}
   \label{J.m2.e:d2Zdvdv} \\
   & \quad
   =
   \frac{\calN}{Z}
   \int \rD \Phi \,
   \Bigl ( \frac{\hbar}{i}  \Bigr )^2 \, 
   \Bigl [ \,
      \Bigl ( \frac{\partial S}{\partial v_i} \Bigr ) \,
      \Bigl ( \frac{\partial S}{\partial v_j} \Bigr )
      -
      \frac{\partial^2 S}{\partial v_i \, \partial v_j} \,
   \Bigr ] \, e^{-S}
   \notag \\
   & \quad
   =
   m^2 \iint [\rd x] \, [\rd x'] \, x_i \, x'_j
   \notag \\
   & \quad
   \times
   \Bigl \{ \,
      g^{\alpha}{}_{\mu} \, g^{\beta}{}_{\nu} \,
      \Expectbig{ \frac{\delta S[\Phi_v]}{\delta \Phi_v^{\alpha}(x)} \, \Phi_v^{\mu}(x) } \,
      \Expectbig{ \frac{\delta S[\Phi_v]}{\delta \Phi_v^{\beta}(x')} \, \Phi_v^{\nu}(x') }
      \notag \\
      & \qquad\quad
      -
      \delta(x,x') \,
      \Expectbig{ \frac{\delta S[\Phi_v]}{\delta \Phi_v^{\alpha}(x)} \, \Phi_v^{\alpha}(x) }
      \notag \\
      & \qquad\quad
      -
      g^{\alpha}{}_{\mu} \, g^{\beta}{}_{\nu} \, 
      \Expectbig{ \frac{\delta^2 S[\Phi_v]}{\delta \Phi_v^{\alpha}(x) \, \delta \Phi_v^{\beta}(x')} \,
      \Phi_v^{\mu}(x) \,  \Phi_v^{\nu}(x') } \,
   \Bigr \} \>.
   \notag
\end{align}
Finally, using the results in Eq.~\eqref{J.m1.e:dZs}, derivatives of the grand potential with respect to $\bv$ is given by
\begin{align}
   \frac{\partial \Omega}{\partial v_i} 
   &=
   - \frac{1}{\beta Z} \, \frac{\partial Z}{\partial v_i} 
   \label{J.m1.e:dOmegas1} \\
   &=
   \frac{i m}{\hbar \beta} 
   \int [\rd x] \, x_i \,  g^{\alpha}{}_{\mu} \, 
   \Expectbig{ \frac{ \delta S[\Phi_v]}{\delta \Phi_v^{\alpha}(x) } \, \Phi_v^{\mu}(x) } \>,
   \notag
\end{align}
and
\begin{align}
   &\frac{\partial^2 \Omega}{\partial v_i \, \partial v_j}
   =
   \beta \, \frac{\partial \Omega}{\partial v_i} \, \frac{\partial \Omega}{\partial v_j}
   -
   \frac{1}{\beta Z} \, \frac{\partial^2 Z}{\partial v_i \, \partial v_j}
   \label{J.m2.e:ddOmegas} \\
   & \quad
   =
   \frac{1}{\beta Z^2} \, \frac{\partial Z}{\partial v_i} \, \frac{\partial Z}{\partial v_j}
   -
   \frac{1}{\beta Z} \, \frac{\partial^2 Z}{\partial v_i \, \partial v_j} \>.
   \notag \\
   & \quad
   =
   -
   \Bigl ( \frac{m^2}{\hbar^2 \beta} \Bigr )
   \biggl \{ \,
      \int [\rd x] \, x_i \, x_j \,
      \Expectbig{ \frac{\delta S[\Phi_v]}{\delta \Phi_v^{\alpha}(x)} \, \Phi_v^{\alpha}(x) } 
      \notag \\
      & \quad
      +
      \iint [\rd x] \, [\rd x'] \, x_i \, x'_j \,
      g^{\alpha}{}_{\mu} \, g^{\beta}{}_{\nu}
      \notag \\
      & \qquad
      \times
      \Expectbig{ \frac{\delta^2 S[\Phi_v]}{\delta \Phi_v^{\alpha}(x) \, \delta \Phi_v^{\beta}(x')} \,
      \Phi_v^{\mu}(x) \,  \Phi_v^{\nu}(x') } \,
   \biggr \} \>.
   \notag
\end{align}
The superfluid density is given by Eq.~\eqref{J.m.e:rhos},
\begin{equation}\label{J.m2.e:rhos-I}
   \rho_s
   =
   \frac{1}{V}
   \Bigl [ \,
      \frac{ \partial^2 \, \Omega(v) }{ \partial v_z^2 } \,
   \Bigr ]_{v=0} \>.   
\end{equation}
So we want to evaluate \eqref{J.m2.e:ddOmegas} at $i = j = z$ and at $\bv = 0$.  Now at zero currents,
\begin{equation}\label{J.m2.e:zerocurrents}
   \frac{\delta S[\Phi_v]}{\delta \Phi_v^{\alpha}(x)}
   =
   J_{\alpha}(x)
   \Rightarrow
   0 \>,
\end{equation}
so the first term in the third line of \eqref{J.m2.e:ddOmegas} vanishes.  So then the superfluid density is given by
\begin{align}
   \rho_s
   &=
   -
   \Bigl ( \frac{m^2}{\hbar^2 \beta V} \Bigr )
   \iint [\rd x] \, [\rd x'] \, z \, z' \, g_{\alpha}{}^{\mu} \, g^{\beta}{}_{\nu}
   \label{J.m2.e:rhos-II} \\
   & \qquad
   \times
   \Expectbig{ \frac{\delta^2 S_{\text{eff}}[\Phi]}{\delta \Phi_{\alpha}(x) \, \delta \Phi^{\beta}(x')} \,
   \Phi_{\mu}(x) \,  \Phi^{\nu}(x') } \>,
   \notag    
\end{align}
where the fields are now to be evaluated at $\bv = 0$.  In leading order we can replace the fields $\Phi^{\alpha}(x)$ by their expectation values so that the superfluid density in \eqref{J.m2.e:rhos-II} becomes
\begin{align}
   \rho_s
   &=
   -
   \Bigl ( \frac{m^2}{\hbar^2 \beta V} \Bigr )
   \iint [\rd x] \, [\rd x'] z \, z' 
   \label{J.m2.e:rhos-III} 
   \bar{\Phi}_{\alpha} \, \calG^{-1,}{}^{\alpha}{}_{\beta}[\Phi](x,x') \, \bar{\Phi}^{\beta} \>,
\end{align}
where $\bar{\Phi}^{\alpha}$ are given in Eq.~\eqref{J.G.e:barPhidefs}.
Expanding $\calG^{-1,}{}^{\alpha}{}_{\beta}(x,x')$ in a Fourier series according to Eq.~\eqref{J.G.e:Gexpand},
and inserting this into \eqref{J.m2.e:rhos-III} and using the fact that the fields are constant, we obtain
\begin{align}
   \rho_s
   \label{J.m2.e:rhos-VI}
   = &
   \Bigl ( \frac{m^2}{2 \hbar^2} \Bigr ) 
   \lim_{q \rightarrow 0}
   \frac{\partial^2}{\partial q_z^2} 
   \Bigl [ \,
      \bar{\Phi}_{\alpha} \, \tilde \calG^{-1,}{}^{\alpha}{}_{\beta}[\Phi](\bq,0) \, \bar{\Phi}^{\beta}
   \Bigr ] \>.
\end{align}
Because the $U(1)$ charge of the auxiliary field $\chi$ is zero, it is convenient to restrict ourselves to the $(1,2,4,5)$ set of indices.  Therefore the relevant part of the inverse Green function $\tilde{G}^{-1,\alpha}{}_{\beta}(0,0)$ is the one we discussed previously in connection with the Goldstone theorem in Sec.~\ref{G.s:goldstone} and is given by the matrix
\begin{equation}\label{J.m2.e:MLOAF-Ix}
   M^{\alpha}{}_{\beta}
   =
   \begin{pmatrix}
      \epsilon_q + \chi' & -A & -\phi^{\ast} & 0 \\
      -A^{\ast} & \epsilon_q + \chi' & 0 & -\phi \\
      -\phi & 0 & \alpha & \beta \\
      0 & -\phi^{\ast} & \beta^{\ast} & \alpha
   \end{pmatrix} \>,
\end{equation}
where we introduced the notations,
\begin{align}\label{J.m2.e:alphabetadefs}
   \alpha
   &=
   \tilde D^{-1,1}{}_{1}[\chi_0](\bq,0)
   =
   \tilde D^{-1,2}{}_{2}[\chi_0](\bq,0)
   \\
   \beta
   &=
   \tilde D^{-1,1}{}_{2}[\chi_0](\bq,0)
   =
   \tilde D^{-1,2}{}_{1}[\chi_0]^{\ast}(\bq,0) \>.
   \notag
\end{align}
with $D^{-1,}{}^{i}{}_{j}[\chi_0](x,x')$ given in Eq.~\eqref{J.ea.e:Dinverse}.  

For small momentum values, the momentum expansion of the Green functions is
\begin{align}
   \alpha &= \alpha_0 + \alpha_1 q^2 + \cdots
   \>,
   \\
   \beta &= \beta_0 + \beta_1 q^2 + \cdots
   \>.
\end{align}
Then, the inverse Green function can be written as
\begin{equation}\label{J.m2.e:Ginvexp}
   \calG^{-1,}{}^{\alpha}{}_{\beta}[\Phi](\bq,0)
   =
   M_0^{\alpha}{}_{\beta}
   +
   M_1^{\alpha}{}_{\beta} \, q^2
   +
   \dotsb \>,
\end{equation}
where
\begin{equation}\label{J.m2.e:M0LOAF-I}
   M_0^{\alpha}{}_{\beta}
   =
   \begin{pmatrix}
      \chi' & -A & -\phi^{\ast} & 0 \\
      -A^{\ast} & \chi' & 0 & -\phi \\
      -\phi & 0 & \alpha_0 & \beta_0 \\
      0 & -\phi^{\ast} & \beta_0^{\ast} & \alpha_0
   \end{pmatrix} \>,
\end{equation}
as before, and 
\begin{equation}\label{J.m2.e:M1LOAF-I}
   M_1^{\alpha}{}_{\beta}
   =
   \begin{pmatrix}
      \hbar^2/2m & 0 & 0 & 0 \\
      0 & \hbar^2/2m & 0 & 0 \\
      0 & 0 & \alpha_1 & \beta_1 \\
      0 & 0 & \beta_1^{\ast} & \alpha_1
   \end{pmatrix} \>.
\end{equation}
We find
\begin{align}
   &\bar{\Phi}_{\alpha} \, M_0^{\alpha}{}_{\beta} \, \Phi^{\beta}
   =
   - 
   2 \, \chi' | \phi |^2 
   + 
   3 \, ( A \, \phi^{\ast\,2} + A^{\ast} \, \phi^2 )
   \label{J.m2.e:PhiM0Phi} \\
   & \qquad
   -
   8 \, \alpha_0 \, | A |^2
   +
   4 \, ( A^2 \, \beta_0^{\ast} + A^{\ast\,2} \, \beta_0 ) \>.
   \notag
\end{align}
and
\begin{align}
   &\bar{\Phi}_{\alpha} \, M_1^{\alpha}{}_{\beta} \, \Phi^{\beta}
   =
   -
   8 \, \alpha_1 \, | A |^2
   +
   4 \, ( A^2 \, \beta_1^{\ast} + A^{\ast\,2} \, \beta_1 ) \>.
\end{align}
We specialize now to the case of real fields and we recall that in LOAF we have (see Eq.~\ref{J.G.e:xxx1})
\begin{align}\label{J.G.e:xxx2}
   \alpha_0 - \beta_0 
   =
   \tilde{D}^{-1,}{}^{1}{}_{1}(0,0)
         -
   \tilde{D}^{-1,}{}^{1}{}_{2}(0,0)
   = 
   \frac{ \phi^2 }{ 2 A } \>.
\end{align}
For $T < T_c$, we have $\phi \ne 0$ and $\chi' = A > 0$. Therefore, in region~I, we obtain
\begin{equation}
   \bar{\Phi}_{\alpha} \, M_0^{\alpha}{}_{\beta} \, \Phi^{\beta}
   =
   4 \, A \phi^2 
   - 
   8 \, A^2 ( \alpha_0 - \beta_0 ) \>,
\end{equation}
which vanishes when we make use of Eq.~\eqref{J.G.e:xxx2}. 
Similarly, $\bar{\Phi}_{\alpha} \, M_0^{\alpha}{}_{\beta} \, \Phi^{\beta}$ vanishes for $T > T_c$, because $\phi = 0$. So, the $M_0^{\alpha}{}_{\beta}$ contribution is identically zero everywhere.
Therefore, we find that the superfluid density is given by
\begin{align}
   \rho_s
   &=
   - \Bigl ( \frac{4 m^2 \, A^2}{\hbar^2} \Bigr ) 
   \lim_{q \rightarrow 0}
   \frac{\partial^2}{\partial q_z^2} 
   \bigl [ \,
      \tilde{D}^{-1,}{}^{1}{}_{1}(\bq,0)
      -
      \tilde{D}^{-1,}{}^{1}{}_{2}(\bq,0)
      \notag \\
      & \qquad \qquad \qquad \qquad
      -
      \tilde{D}^{-1,}{}^{2}{}_{1}(\bq,0)
      +
      \tilde{D}^{-1,}{}^{2}{}_{2}(\bq,0) \,
   \bigr ]
   \notag \\
   &=
   - \Bigl ( \frac{8 m^2 \, A^2}{\hbar^2} \Bigr ) 
   \lim_{q \rightarrow 0}
   \frac{\partial^2}{\partial q_z^2} 
      \bigl [ \,
         \tilde{D}^{-1,}{}^{1}{}_{1}(\bq,0)
         -
         \tilde{D}^{-1,}{}^{1}{}_{2}(\bq,0)
      \bigr ]
   \>.
\end{align}
In leading order, the superfluid density is
\begin{equation}\label{J.m2.e:rhos-calcI}
   \rho_s
   =
   - \Bigl ( \frac{4 m^2 \, A^2}{\hbar^2} \Bigr ) \, 
      4 \, (\alpha_1 - \beta_1) 
   + 
   \dotsb \>.
\end{equation}
In order to write \eqref{J.m2.e:rhos-calcI} in terms of a component of the Green function, we compute the inverse of Eq.~\eqref{J.m2.e:MLOAF-Ix} and select the $M^{-1,4}{}_{4}$ component.  For real fields, this gives
\begin{equation}\label{J.m2.e:M44inverse}
   \frac{1}{M^{-1,4}{}_{4}(q)}
   =
   \frac{( \alpha^2 - \beta^2 ) ( \chi^{\prime\,2} - A^2 ) - 2 ( \alpha - \beta ) \phi^2 A }
        {\alpha ( \chi^{\prime\,2} - A^2 ) - \chi' \, \phi^2 } \>.
\end{equation}
For $T < T_c$ where $\chi' = A$ and when $T > T_c$ where $\phi = 0$, \eqref{J.m2.e:M44inverse} reduces to $ 2 \, ( \alpha_1 - \beta_1 ) \, q^2$ in both cases.  So we find that
\begin{equation}\label{J.m2.e:M44inverse-II}
   \lim_{q \rightarrow 0} \frac{1}{ q^2 M^{-1,4}{}_{4}(q)} 
   =
   2 \, ( \alpha_1 - \beta_1 ) \>,
\end{equation}
so that \eqref{J.m2.e:rhos-calcI} can be written as
\begin{equation}\label{J.m2.e:rhos-calcII}
   \rho_s
   =
   -\Bigl ( \frac{8 m^2 \, A^2}{\hbar^2} \Bigr ) \, 
   \lim_{q \rightarrow 0} \frac{1}{ q^2 M^{-1,4}{}_{4}(q)} \>.
\end{equation}
A somewhat tedious but straightforward calculation shows that by inverting the full $5 \times 5$ inverse Green function given in Eq.~\eqref{J.G.e:calGinv-I}, we also obtain
\begin{equation}\label{J.m2.e:calG44comp}
   \frac{1}{\tilde{\calG}^{A}{}_{A}(\bq,0)}
   =
   2 \, ( \alpha_1 - \beta_1 ) \, q^2 + \dotsb \>,
\end{equation}
for both $T < T_c$ and $T > T_c$.  So, Eq.~\eqref{J.m2.e:rhos-calcII} can be written as
\begin{equation}\label{J.m2.e:rhos-calcIII}
   \rho_s
   =
   - \frac{8 m^2 \, A^2}{\hbar^2} \, 
   \lim_{q \rightarrow 0} \frac{1}{ q^2 \tilde{\calG}^{A}{}_{A}(q,0) } \>,
\end{equation}
which is the Josephson relation.  We note that the existence of the pole in $\tilde{\calG}^{A}{}_{A}(q,0)$, which is guaranteed by Goldstone's theorem when the $U(1)$ symmetry is broken, is what leads to nonzero superfluid density.

%
%
\section{\label{J.s:conclusions}Conclusions}

To summarize, in this paper we derived the Josephson relation for a dilute Bose gas in the framework of an auxiliary-field resummation of the theory in terms of the normal- and anomalous-density condensates. In the self-consistent mean-field approximation to our auxiliary-filed formalism, the phase diagram of this theory features two critical temperatures, $T_c < T^\star$, associated with the presence in the system of the BEC condensate and superfluid state, respectively. For temperatures below $T_c$, a BEC condensate is present in the system, whereas $T^\star$ signals the onset of superfluidity in the system. As such, for all dilute systems of interacting Bose gases, the LOAF approximation predicts the possibility of a superfluid state in the absence of a BEC condensate in the temperature range between $T_c$ and $T^\star$. The density of the superfluid state is controlled by a second order parameter. The Josephson relation identifies this second order parameter as the square of the anomalous-density condensate. This result contradicts the usual Bose gas theory that does not feature an anomalous condensate, which predicts that the superfluid density is proportional to the BEC condensate density. However, our theory is consistent with the case of dilute Fermi gases, where the BCS theory shows that the superfluid density is proportional to the square of the gap parameter. In this sense, the auxiliary-field formalism discussed here provides a unified approach to the study of fermonic and bosonic atom gases. 

Whereas the Goldstone theorem is a statement about the vanishing of the momentum-independent part of the determinant of the full inverse Green function of the theory, the Josephson relation is related to the coefficient of the determinant proportional to the square of the momentum. The Josephson relation for the superfluid density reads
\begin{equation}
   \rho_s = 
       - \,
   \frac{8 m^2 \, A^2}{\hbar^2} \,
   \lim_{q \rightarrow 0}
   \frac{1}{q^2 \, \tilde{\calG}^{A}{}_A(q,0)} 
   \>,
\end{equation}
where $\tilde{G}^{A}{}_A(q,0)$ is the propagator corresponding to the expectation value $\Expect{A^\star A}$, evaluated for zero energy transfer. 
These results will remain preserved at all orders in the many-body approximations of the auxiliary-field formalism.


\acknowledgements

Work performed in part under the auspices of the U.S. Department of Energy.
The authors would like to thank the Santa Fe Institute for its hospitality during this work.

\newpage
%
%
\bibliography{johns}
%
%
\end{document}